\documentclass[%
 reprint,
 amsmath,amssymb,
 aps,
]{revtex4-2}
\usepackage{braket}
\usepackage{graphicx}
\usepackage{dcolumn}
\usepackage{bm}
\usepackage{appendix}
\usepackage{color}
\usepackage{mathtools}
\usepackage{comment}

\newcommand{\1}{\mbox{1}\hspace{-0.25em}\mbox{l}}

\begin{document}

\preprint{APS/123-QED}

\title{Universal quantum computation using quantum annealing with the transverse-field Ising Hamiltonian}

\author{Takashi Imoto}
\email{takashi.imoto@aist.go.jp}
\affiliation{Research Center for Emerging Computing Technologies, National Institute of Advanced Industrial Science and Technology (AIST), 1-1-1 Umezono, Tsukuba, Ibaraki 305-8568, Japan.}

\author{Yuki Susa}  
\affiliation{Secure System Platform Research Laboratories, NEC Corporation, Kawasaki, Kanagawa 211-8666, Japan}
\affiliation{NEC-AIST Quantum Technology Cooperative Research Laboratory, National Institute of Advanced Industrial Science and Technology (AIST), Tsukuba, Ibaraki 305-8568, Japan}

\author{Ryoji Miyazaki}  
\affiliation{Secure System Platform Research Laboratories, NEC Corporation, Kawasaki, Kanagawa 211-8666, Japan}
\affiliation{NEC-AIST Quantum Technology Cooperative Research Laboratory, National Institute of Advanced Industrial Science and Technology (AIST), Tsukuba, Ibaraki 305-8568, Japan}

\author{Yuichiro Matsuzaki}
\affiliation{
Department of Electrical, Electronic, and Communication Engineering, Faculty of Science and Engineering, Chuo University, 1-13-27 Kasuga, Bunkyo-ku, Tokyo 112-8551, Japan.
}
\date{\today}

\begin{abstract}
Quantum computation is a promising emerging technology, and by utilizing the principles of quantum mechanics, it is expected to achieve faster computations than classical computers for specific problems.
There are two distinct architectures for quantum computation: gate-based quantum computers and quantum annealing.
In gate-based quantum computation, we implement a sequence of quantum gates that manipulate qubits. This approach allows us to perform universal quantum computation,
yet they pose significant experimental challenges for large-scale integration.
On the other hand, with quantum annealing, the solution of the optimization problem can be obtained by preparing the ground state.
Conventional quantum annealing devices with transverse-field Ising Hamiltonian, such as those manufactured by D-Wave Inc., achieving around 5000 qubits, are relatively more amenable to large-scale integration but are limited to specific computations.
In this paper, we present a practical method for implementing universal quantum computation within the conventional quantum annealing architecture using the transverse-field Ising Hamiltonian. Our innovative approach relies on an adiabatic transformation of the Hamiltonian, changing from transverse fields to a ferromagnetic interaction regime, where the ground states become degenerate. Notably, our proposal is compatible with D-Wave devices, opening up possibilities for realizing large-scale gate-based quantum computers.
This research bridges the gap between conventional quantum annealing and gate-based quantum computation, offering a promising path toward the development of scalable quantum computing platforms.

\begin{description}
\item[Usage]
Secondary publications and information retrieval purposes.
\item[Structure]
You may use the \texttt{description} environment to structure your abstract;
use the optional argument of the \verb+\item+ command to give the category of each item. 
\end{description}
\end{abstract}

\maketitle




\section{Introduction}
Quantum computers, utilizing the principles of quantum mechanics, are expected to provide faster computation than classical computers.
In the field of quantum computation, two distinctive architectures have emerged: gate-based quantum computers and quantum annealing, each offering unique approaches to harnessing quantum phenomena
\cite{kadowaki1998quantum, farhi2000quantum, farhi2001quantum}.

The gate-based quantum computer consists of sequential unitary quantum gates, enabling universal quantum computation.
In particular, we remark that arbitrary unitary quantum gates can be constructed by combining several types of unitary gates, called universal gate sets\cite{barenco1995elementary, boykin2000new, kitaev1995quantum, shi2002both}.
Typically, either phase accumulation (such as Ramsey interferometry) or application of an oscillating field (such as Rabi oscillations) is adopted to implement the gate operations for a solid-state quantum computer.
It has the potential for a wide array of applications, including quantum chemistry, machine learning, financial engineering, optimization problems, and decrypting ciphers\cite{shor1994algorithms, shor1999polynomial,harrow2009quantum, kitaev1995quantum, nielsen2010quantum, grover1996fast}.
Most quantum algorithms to achieve quantum speed-up need quantum error collection, exemplified by a surface code\cite{kitaev1997quantum}.
This paradigm, known as fault-tolerant quantum computation(FTQC), requires thousands of physical qubits per logical qubit to effectively suppress the noise effect.
Consequently, it is estimated that FTQC will at least require 10 million qubits to perform computations beyond the capabilities of existing classical computers.
Nevertheless, the scalability of gate-based quantum computers poses substantial experimental challenges, limiting them to utilizing only around a hundred qubits with current technology.\\

In contrast to gate-based quantum computation, quantum annealing(QA) offers an alternative method, focused on solving combinatorial optimization problems by preparing the ground state of an Ising Hamiltonian.
Quantum annealers with transverse-field Ising Hamiltonians, such as those developed by D-Wave Systems, have made thousands of qubits available for these tasks\cite{boixo2014evidence, johnson2011quantum}.
QA has many applications besides solving combinatorial optimization problems, such as machine learning\cite{adachi2015application,sasdelli2021quantum, date2021adiabatic, wilson2021quantum, kumar2018quantum, kurihara2014quantum, willsch2020support, heidari2023quantum, woun2023adiabatic, neven2009training, bosch2023neural, pudenz2013quantum} and physical simulation\cite{zhou2021experimental, kairys2020simulating, harris2018phase, king2018observation, king2023quantum}.
QA can perform adiabatic time evolution for a sufficiently long time, and then find the ground state of the Hamiltonian with high accuracy. 
Also, the adiabaticity of this process is guaranteed by the adiabatic theorem\cite{kato1950adiabatic,messiah2014quantum,jansen2007bounds}.
Significant efforts have been devoted to achieving adiabaticity in a short time.

Adiabatic quantum computation is an approach to implement universal quantum computation with adiabatic dynamics.
\cite{aharonov2008adiabatic, biamonte2008realizable}.
However, adiabatic quantum computation encounters significant challenges.
One primary obstacle is the requirement for 
many ancillary qubits.
As the corresponding circuit depth increases, the necessary number of ancillary qubits also increases.
Additionally, it necessitates the use of the so-called non-stoquastic Hamiltonian, which is experimentally difficult to realize. 
The demonstration with the non-stoquastic Hamiltonian was limited to a few qubits.



Various techniques for qubit realization have been proposed, with superconducting qubits emerging as a particularly promising approach. 
Within this domain, transmons and flux qubits\cite{mooij1999josephson, orlando1999superconducting, clarke2008superconducting,makhlin2001quantum} are known as the primary methods for realizing qubits using superconductivity. 
Flux qubits, which are used in D-Wave devices, are renowned for their scalability without the need for introducing microwaves\cite{johnson2011quantum,boixo2014evidence}.
However, their application is limited to Quantum annealing(QA) as they cannot perform all universal gate operations.
In contrast, transmon qubits despite their integration challenges, are capable of realizing a universal gate set with high fidelity.
This capability was notably demonstrated in the experiment conducted by Google Inc. for achieving quantum supremacy.
Recently, this scheme using cold atoms in qubit construction has attracted a great deal of attention\cite{levine2019parallel,chew2022ultrafast,bluvstein2023logical}.
This approach is anticipated to have high precision gate operations and long coherence time.

\begin{figure*}
    \centering
    \includegraphics[width=1.02\linewidth]{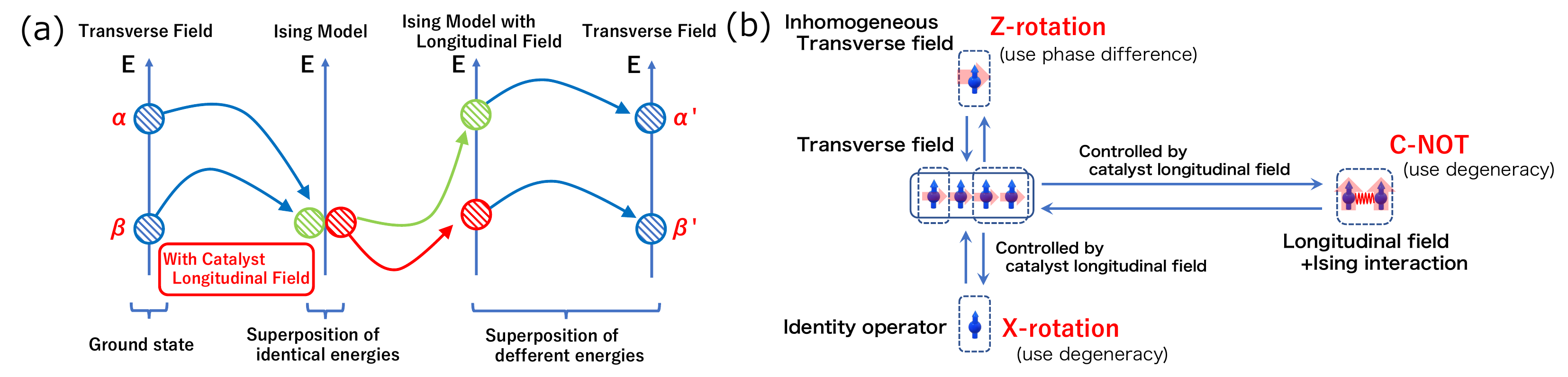}
    \caption{(a)Illustration of our method to implement an $X$ rotation gate. 
    First, we prepare a superposition between
    the ground and first excited states with the transverse field Hamiltonian.
    The process initiates with the preparation of an all-plus state, which serves as the ground state of the transverse field.
    This state is the starting point for implementing quantum annealing with a degenerate Ising model.
    Second, we decrease the amplitude of the transverse fields while we increase the amplitude of the Ising interaction and longitudinal field. Third, we let the Hamiltonian have degenerate ground states. Fourth, we increase the amplitude of the transverse fields while we decrease the amplitude of the Ising interaction and longitudinal field. Finally, the population of the ground state and first excited state is changed when the Ising interaction and longitudinal field vanishes.
    (b)Schematic of Gate-Based Quantum Computation Construction.
    This figure illustrates the comprehensive framework for constructing gate-based quantum computation using our proposed scheme.
    We regard the center of this figure which is under the transverse magnetic field as the idling state.
    The X-rotation gate and the controlled-not gate are implemented by the procedures of the section 
    \ref{sec:propose_c-not} and the section \ref{sec:realize_x_rotation}.
    On the other hand, the Z-rotation gate is realized in the method described in Section \ref{sec:z-axis_op} using the difference of phase.
    Hence, we can realize the gate-based quantum computation using the transverse Ising model.
    }
    \label{fig:concept1}
\end{figure*}

In this paper, we present a practice method for implementing universal quantum computation within the conventional quantum annealing architecture using the transverse-field Ising Hamiltonian.
Our method relies on an adiabatic transformation of the Hamiltonian, changing from transverse fields to a ferromagnetic interaction regime where the ground states become degenerate.
Importantly, the degeneracy provides a way to prepare not only the ground state but also the excited states, which makes it possible to implement the X-gate rotation and controlled-X rotation gate.
Unlike the previous approach to realize a solid-state quantum computer, we do not use either the phase accumulation or the application of oscillating fields to implement the X-gate rotation and controlled-X rotation gate. 
A notable advantage of our method is its independence from microwave pulses, enhancing scalability potential in integrated systems.
Moreover, our proposal is compatible with D-Wave devices, opening up prospects for realizing large-scale gate-based quantum computers.
This research bridges the gap between quantum annealing and gate-based quantum computation, paving the way for a scalable quantum computing platform.

The rest of this paper is organized as follows.
Section \ref{sec:comprehensive_review} is devoted to a comprehensive review of quantum annealing and universal quantum computation.
In Section \ref{sec:superposition}, we elucidate the method for adiabatically creating superposition states through degeneracy.
In Section \ref{sec:realization_of_gate_op}, we introduce gate operation using our scheme. 
Section \ref{sec:numerical_simulation} describes numerical simulations conducted to evaluate the performance of our scheme. 
In Section \ref{sec:experimental_demonstration}, we perform an experimental demonstration of several gate operations using a D-Wave device.
The construction of a gate-based quantum computer using a D-Wave device is
Section \ref{sec:construction_using_d-wave} will construct a gate-based quantum computer using a D-Wave device.
We conclude with Section \ref{sec:conclusion}.

\section{Review of quantum annealing}\label{sec:comprehensive_review}





Let us review quantum annealing.
A solution of the combinatorial optimization problems can be embedded into a ground state of an Ising Hamiltonian $H_{P}$. 
We can use quantum annealing to find the ground state of an Ising Hamiltonian via adiabatic dynamics.
Let us consider a trivial Hamiltonian $H_{D}$, such as transverse fields called the driver Hamiltonian.
We define the annealing Hamiltonian as
\begin{align}
    H(t)\equiv \biggl(1-\frac{t}{T}\biggr)H_{D}+\frac{t}{T}H_{P}.\label{eq:ann_ham}
\end{align}
where $T$ denotes the annealing time.
We prepare the ground state of the drive Hamiltonian and let the state evolve by the annealing Hamiltonian in Eq \eqref{eq:ann_ham} from $t=0$ to $t=T$.
A notable aspect of QA is that when an annealing Hamiltonian maintains a certain symmetry, the dynamics is constrained within the same sector by block-diagonalization and QA could fail
\cite{farhi2000quantum,imoto2022obtaining,imoto2022quantum, hatomura2022quantum, matsuzaki2022generation}.
Unless such symmetry exists, 
the quantum adiabatic theorem guarantees that
we obtain the ground state of the problem Hamiltonian for a long $T$.

\section{Realization of gate operation}\label{sec:realization_of_gate_op}
In this section, we explain how to implement the single-qubit rotation gate and the controlled-not gate by quantum annealing. 
In our method, when we do not perform any gates, we set the transverse magnetic field Hamiltonian, which we call the idling Hamiltonian.
To perform the Z rotation gate, we use the phase accumulation between the ground state and the first excited states. On the other hand, to perform either X-gate rotation or controlled not gate, we adopt a unique approach.
Our method consists of two key processes: quantum annealing with a degenerate problem Hamiltonian and reverse annealing.
We control the gate operation by adjusting the longitudinal magnetic field during the quantum annealing.

\subsection{Change in the amplitude of the superposition of the energy eigenstate of a degenerate Hamiltonian
}

We describe the method for controlling the 
amplitude of the superposition of energy eigenstates of the degenerate Hamiltonian.
We focus on scenarios where the ground states of the problem Hamiltonian $H_{P}$ are doubly degenerate.
The two ground states of the problem Hamiltinian $H_{P}$ are denoted by $\ket{\phi_{1}},\ \ket{\phi_{2}}$.
After QA with a sufficiently long $T$, we obtain the following state
\begin{align}
    \ket{\Phi}=\alpha\ket{\phi_{1}}+\sqrt{1-\alpha^{2}}\ket{\phi_{2}}
\end{align}
where $\alpha$ is the amplitude of $\ket{\phi_{1}}$.
Importantly, $\alpha$ will be determined by the process of QA.

First, to control the amplitude $\alpha$, we introduce the following "forward" Hamiltonian
\begin{align}
    H_{F}(t)= 
    \begin{dcases}
    \Bigl(1-\frac{t}{T}\Bigr)H_{D}+\frac{t}{T}H_{P}+\frac{t}{T}h_{z} H_{C}, &  \Bigl(t<\frac{T}{2}\Bigr) \\
    \Bigl(1-\frac{t}{T}\Bigr)H_{D}+\frac{t}{T}H_{P}+\Bigl(1-\frac{t}{T}\Bigr)h_{z} H_{C}, &  \Bigl(t>\frac{T}{2}\Bigr)
  \end{dcases}\label{eq:parames_ann_ham}
\end{align}
where $H_{C}$ is the catalytic Hamiltonian and $h_{z}$ is the amplitude parameter of the catalytic term.
We should remark that when $h_{z}=0$, this Hamiltonian is identical to the conventional annealing Hamiltonian.
Also,
the properties $H(t=0)=H_{D}$ and $H(t=T)=H_{P}$ hold independent on amplitude parameter $h_{z}$.
We can control the population $\alpha$ in (\ref{eq:parames_ann_ham}) via tuning the amplitude parameter $h_{z}$ in (\ref{eq:parames_ann_ham}).
Throughout this paper, we choose the longitudinal magnetic field
\begin{align}
    H_{C}=\sum_{i}^{L}\eta_{i}\hat{\sigma}_{i}^{(z)}
\end{align}
as the catalytic Hamiltonian $H_{C}$.
We call this procedure which is the control of the amplitude for eigenstates for the forward part.
Several superpositions of computational basis are prepared using our scheme.
We discuss how to implement specific superpositions in detail in the Appendix \ref{sec:superposition}.

Next, we introduce a modified problem Hamiltonian.
To resolve the degeneracy of the problem Hamiltonian, we add the longitudinal magnetic field.
We call this the modified problem Hamiltonian, which is expressed by $\Tilde{H}_{P}$.
As we will explain later, the specific form of $\Tilde{H}_{P}$ depends on the gate to be implemented.

Finally, we introduce the reverse part to convert the problem Hamiltonian to the drive Hamiltonian.
The reverse Hamiltonian 
is as follows.
\begin{align}
    H_{R}(t)=\frac{t}{T}H_{D}+\biggl(1-\frac{t}{T}\biggr)\tilde{H}_{P}\label{eq:reverse_ann_ham}
\end{align}
We note that
we have $H_{R}(t=0)=\tilde{H}_{P}$ and $H_{R}(t=T)=H_{D}$.
Thus, the Hamiltonian after the reverse part is a drive Hamiltonian.

\begin{figure}
    \centering
    \includegraphics[width=0.8\linewidth]{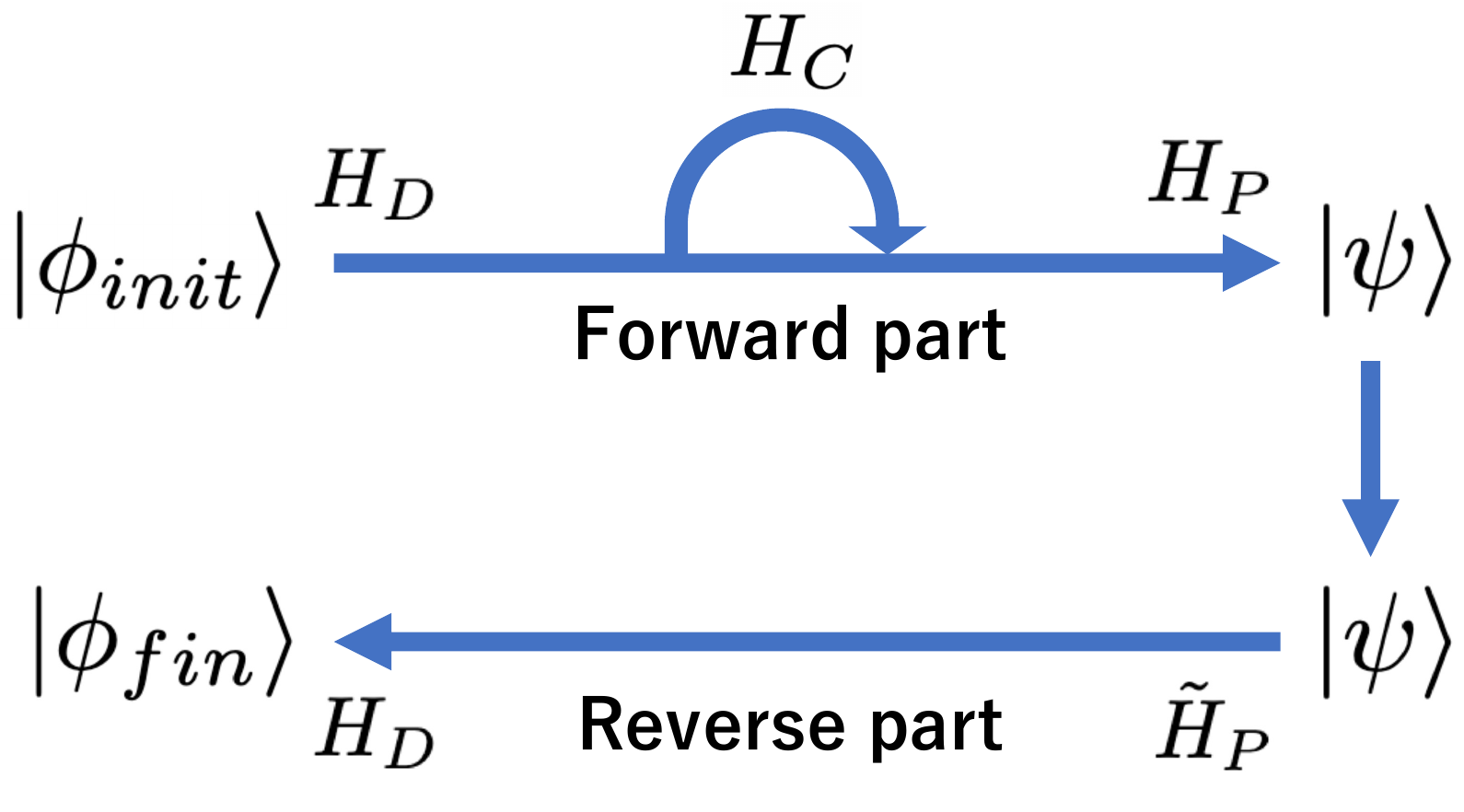}
    \caption{Schematic figure of the Forward and Reverse Parts.
    In the forward part, our primary objective is to control the amplitude of
    the degenerated ground states precisely when we finish the forward part.
    Following the completion of the forward part, the next crucial step involves the addition of a longitudinal magnetic field to the degenerate Ising model.
    This addition is strategically implemented to resolve the degeneracy that was initially present in the ground states. 
    Finally, we change the Hamiltonian to the transverse magnetic field adiabatically.
    }
    \label{fig:forward_reverse_concept}
\end{figure}

In short, our scheme consists of two parts: The forward part and the reverse part.
We use the forward part to control the amplitude $\alpha$ using the Hamiltonian in (\ref{eq:parames_ann_ham}).
After the forward part, the Hamiltonian is the Ising model and the eigenstates of this Hamiltonian are represented by the computational basis.
On the other hand, the reverse part is the operation to return to the drive (idling) Hamiltonian 
The concept of the forward part and the reverse part is illustrated in Fig.\ref{fig:forward_reverse_concept}.

\subsection{X-rotation gate
}\label{sec:realize_x_rotation}

To implement the X-rotation gate, the drive Hamiltonian $H_{D}$ is set as the transverse magnetic field as follows
\begin{align}
    H_{D}=-\hat{\sigma}^{(x)}\label{eq:x-axis_drive}
\end{align}
In addition, we choose 
\begin{align}
    H_{P}=-\1\label{eq:x-axis}
\end{align}
as the problem Hamiltonian. 
We remark that $\1$ is represented as the identity operator.
We note that the Hamiltonian (\ref{eq:x-axis}) has degenerated two ground states:$\ket{\uparrow}$ and $\ket{\downarrow}$.
Thus, after the forward part, we can control the amplitude of the two ground states.
On the other hand, we set the reverse problem Hamiltonian to
\begin{align}
    \tilde{H}_{P}=\hat{\sigma}^{(z)}\label{eq:x-axis_reverse}.
\end{align}
Here, the ground state is $|\downarrow \rangle $ while the first excited state is $|\uparrow \rangle $.
After the reverse process, $|\downarrow \rangle $ is changed into $|-\rangle $ while
$|\uparrow \rangle $ is changed into $|+\rangle $
as long as 
the adiabatic condition is satisfied.
Therefore, the above process makes the amplitude of each eigenstate $\ket{+}$, $\ket{-}$ controllable using the amplitude parameter $h_{z}$.
We discuss this point in the Appendix \ref{xrotationappendix}.

We need to adjust a relative phase after we perform the X-rotation gate.
For a given initial state of $\alpha |+\rangle + \beta e^{i\theta }|-\rangle $,
we obtain $\alpha' |+\rangle + \beta' e^{i(\theta +\theta')}|-\rangle $
after performing the X-rotation gate
where $\alpha$, $\beta$, $\alpha'$, $\beta'$, $\theta$, and $\theta'$ are real coefficients. We can control the amplitude of $\alpha'$ and $\beta'$ by controlling the scheduling of $H_C$. On the other hand, we need to apply a Z-rotation gate to change the relative phase from $e^{i(\theta +\theta')}$ to $e^{i\theta }$. 
We can estimate the value of $\theta'$ using either numerical simulations or actual experiments. 
We will explain how to implement the Z-rotation gate later.

\subsection{Controlled-not gate}\label{sec:propose_c-not}

\begin{figure}
    \centering
    \includegraphics[width=0.75\linewidth]{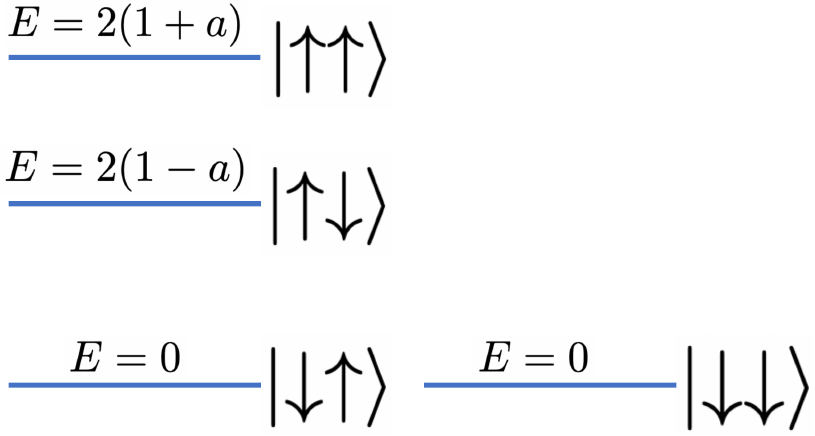}
    \caption{Energy diagram of
    the problem Hamiltonian for the controlled-not gate.
    The constant $a$ satisfies the condition $0<a<1$.
    This figure shows that the $\ket{\downarrow\uparrow}$ and $\ket{\downarrow\downarrow}$ are degenerated ground states and $\ket{\uparrow\uparrow}$ and $\ket{\uparrow\downarrow}$
    are non-degenerate excited states.}
    \label{fig:cnot1_1logical_1phys_enegy}
\end{figure}

To implement the controlled-not gate operation by using our approach, the drive Hamiltonian is set as the transverse magnetic field as follows.
\begin{align}
    H_{D}=-\hat{\sigma}_{1}^{(x)}-\frac{1}{2}\hat{\sigma}_{2}^{(x)}\label{eq:c-not_drive_ham}
\end{align}
The system described in the Hamiltonian (\ref{eq:c-not_drive_ham}) does not have any degeneracy.
Also, we choose 
\begin{align}
H_{P}=\Bigl(\hat{\sigma}_{1}^{(z)}+\1\Bigr)\Bigl(a\hat{\sigma}_{2}^{(z)}+\1\Bigr)\label{eq:c-not_prob_ham}
\end{align}
as the problem Hamiltonian, where the parameter a is satisfied with the inequality $0<a<1$.
We note that $\ket{\downarrow\downarrow}$ and $\ket{\downarrow\uparrow}$ are degenerated ground states.
In addition, $\ket{\uparrow\uparrow}$ and $\ket{\uparrow\downarrow}$ are 
non-degenerate eigenstates.
Thus, the energy spectrum of this problem Hamiltonian is illustrated in Fig \ref{fig:cnot1_1logical_1phys_enegy}.

\begin{figure}
    \centering
    \includegraphics[width=0.55\linewidth]{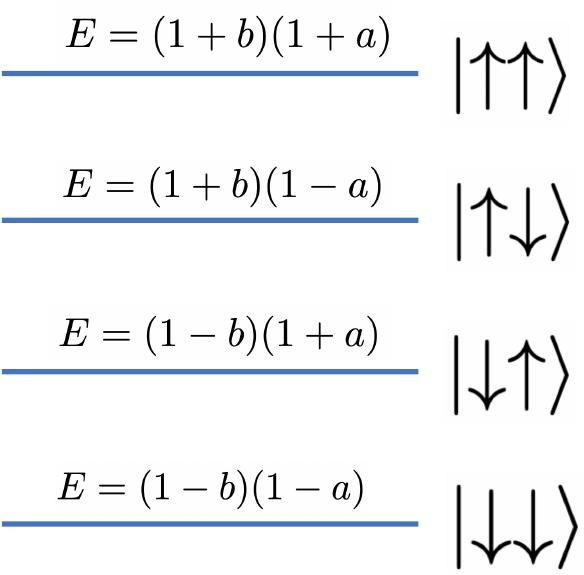}
    \caption{Energy diagram of
    the problem Hamiltonian $\tilde{H}_{P}$ (\ref{eq:c-not_return_prob_ham}) for the reverse process when we perform a controlled-not gate. 
    For this system, there is no degeneracy.
    }
    \label{fig:energy_cnot_reverse_prob_ham}
\end{figure}

To implement the reverse part, we choose the reverse problem Hamiltonian as
\begin{align}
    \tilde{H}_{P}=\Bigl(b\hat{\sigma}_{1}^{(z)}+\1\Bigr)\Bigl(a\hat{\sigma}_{2}^{(z)}+\1\Bigr)\label{eq:c-not_return_prob_ham}.
\end{align}
We assume that $a<b$ and $0<b<1$ are satisfied. In this case, 
the order of the energy eigenstates remains the same as that in Fig \ref{fig:energy_cnot_reverse_prob_ham}.
After the reverse process, $|\downarrow \downarrow \rangle $, $|\downarrow \uparrow \rangle $, $|\uparrow \downarrow \rangle $, and $|\uparrow \uparrow \rangle $ are changed into $|- -\rangle $, $|- +\rangle $, $|+ -\rangle $, and $|++\rangle $, respectively
as long as 
the adiabatic condition is satisfied.
Similar to the case of the X-rotation gate, we need to adjust the relative phase after performing the controlled-not gate.
Here, we assume that the interaction between qubits is off when we apply only the transverse magnetic field. Also, we assume that the interaction gradually increases as we decrease the transverse magnetic fields. This assumption is valid for the D-wave cloud system. Due to this assumption, while we perform the controlled-not gate between two qubits, the other qubits are unaffected by the operation.

\subsection{Z-axis rotation
}\label{sec:z-axis_op}

We use the longitudinal magnetic field when we implement the Z-axis gate operation.
Specifically, by applying the longitudinal magnetic field, an energy difference is given to the two eigenstates, and 
we can implement the Z-axis rotation by phase difference with the time evolution as follows.
\begin{align}
e^{i\hat{\sigma}^{(z)}t}&\Bigl(\alpha\ket{\uparrow}+\beta\ket{\downarrow}\Bigr)\notag\\
&=\alpha e^{it}\ket{\uparrow}+\beta e^{-it}\ket{\downarrow}.
\end{align}
where $\alpha\ket{\uparrow}+\beta\ket{\downarrow}$ is initial state.
This method is similar to the Ramsey interferometry.

\subsection{Interpretation of the Forward part and Reverse part using Hadamard gate}
\begin{figure}
    \centering
    \includegraphics[width=1.0\linewidth]{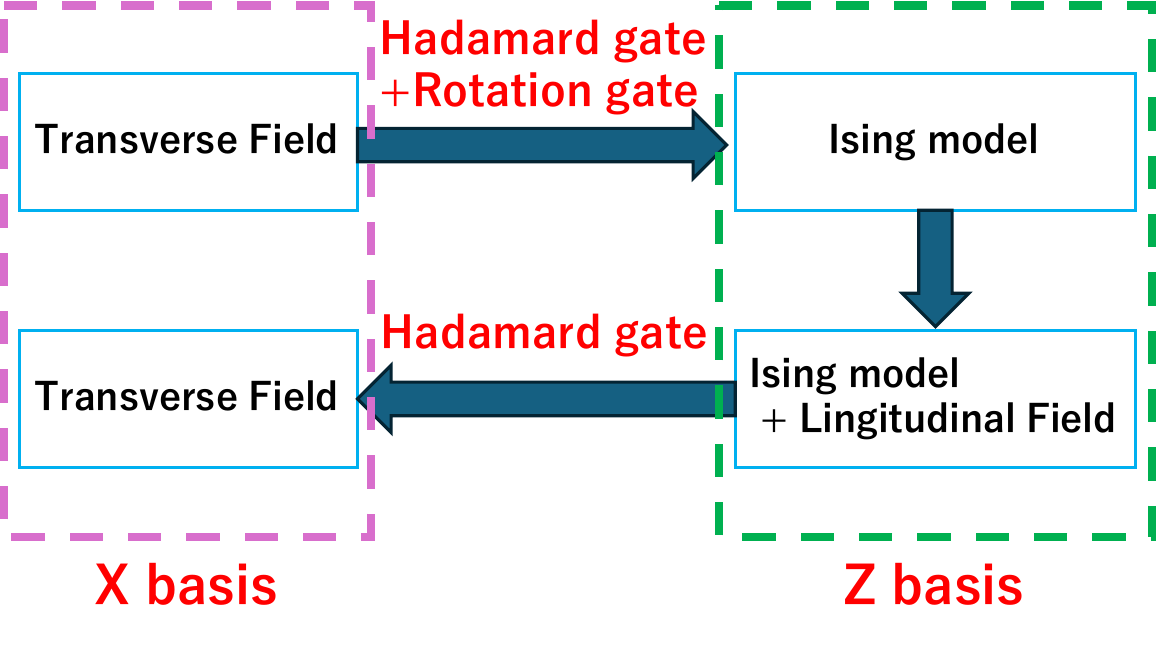}
    \caption{Illustration of our approach using Hadamard gate from the perspective of X-basis and Z-basis.
    }  
    \label{fig:Illustration of our approach using Hadamard gate}
\end{figure}

Our scheme involves two adiabatic dynamics, referred to as the forward and reverse parts, which are associated with both X-axis rotation and the controlled-NOT gate operation. These two dynamics can be interpreted as basis transformations and changing amplitude of each eigenstate to elucidate their roles within the scheme.
(Strictly speaking, these two dynamics occur simultaneously when the state degenerates. However, for the sake of convention, we separately explain these.)
The forward part corresponds to a transformation from the X-basis to the Z-basis. Specifically, this transformation can be expressed as: 
\begin{align} \alpha\ket{+} + \beta\ket{-} \rightarrow \alpha\ket{\uparrow} + \beta\ket{\downarrow}, 
\end{align} 
changing amplitude of each eigenstate within the Z-basis: 
\begin{align} \alpha\ket{\uparrow} + \beta\ket{\downarrow} 
\rightarrow \alpha'\ket{\uparrow} + \beta'\ket{\downarrow}. 
\end{align}
Similarly, the reverse part corresponds to a transformation from the Z-basis back to the X-basis represented as
\begin{align}
    \alpha'\ket{\uparrow}+\beta'\ket{\downarrow}\rightarrow\alpha'\ket{+}+\beta'\ket{-}
\end{align}
This process is illustrated in Fig.\ref{fig:Illustration of our approach using Hadamard gate} which demonstrates our approach using Hadamard gate from the perspective of X-basis and Z-basis.

\section{Numerical Simulation}\label{sec:numerical_simulation}

\begin{figure*}[t]
    \centering
    \includegraphics[width=0.75\linewidth]{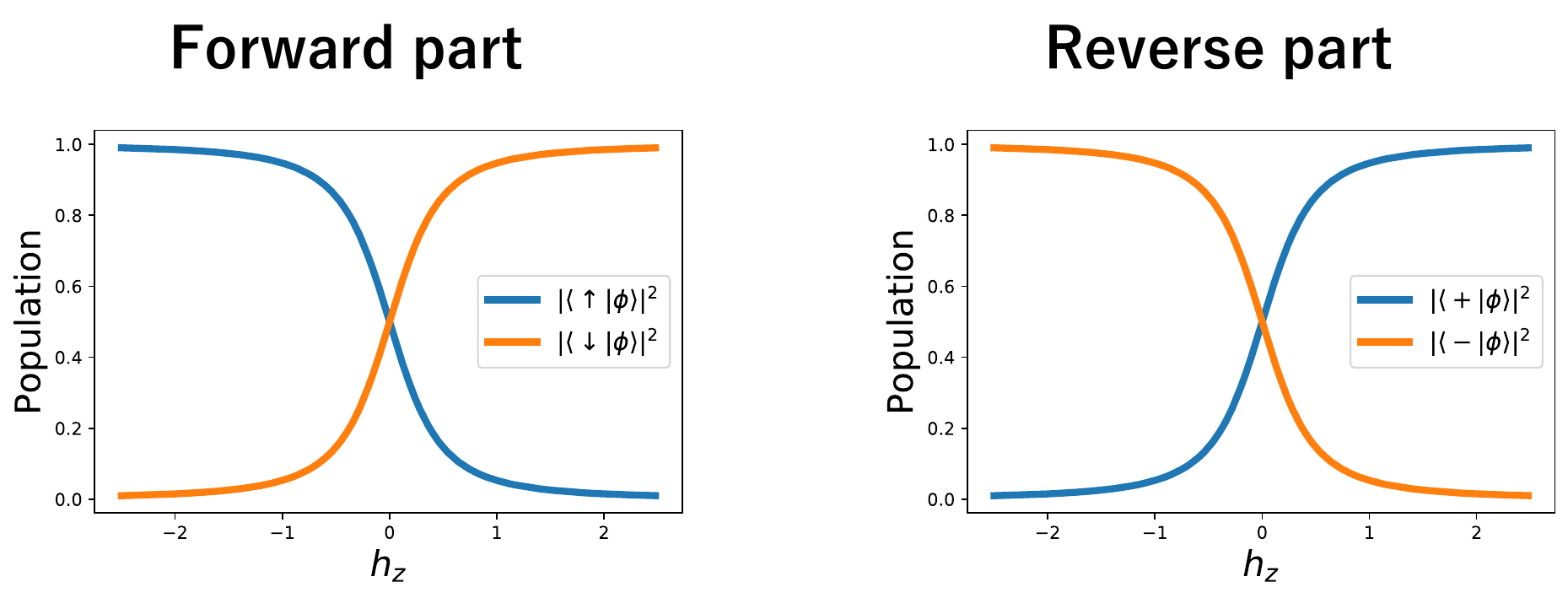}
    \caption{
    Numerical results of the X-rotation gate with our method.
    We plot the population
    of $\ket{\uparrow}$ and $\ket{\downarrow}$ after the forward part of our method. 
    Also, we plot the population of $\ket{-}$ and $\ket{+}$ after the reverse part.
    After the reverse part, the result shows that we can control the population of $\ket{\uparrow}$ and $\ket{\downarrow}$ by tuning the longitudinal magnetic field $h_z$.
    These results illustrate that we can implement the X-rotation gate with any angle with our method.
    }
    \label{fig:forward_reverse_population_2000}
\end{figure*}

In this section, 
we perform numerical simulations of our method
to implement the X rotation and controlled-not gates.

\subsection{X-axis gate}\label{sec:X-axis_gate_numerical_sinulation}

In this subsection, 
we show the results of the numerical simulations to implement the X-rotation gate of our method.
The Hamiltonian $H_{D}$, $H_{P}$, and $\tilde{H}_{P}$ are choosen by (\ref{eq:x-axis_drive}), (\ref{eq:x-axis}), and (\ref{eq:x-axis_reverse}),  respectively.
We adopt the forward Hamiltonian (\ref{eq:reverse_ann_ham}) and the reverse Hamiltonian (\ref{eq:parames_ann_ham}) to implement this gate operation.
In our simulation, the initial state is set to $\ket{+}$ and the total annealing time $T$ is chosen as $2000$.

We plot the population after the forward part against $h_{z}$ in Fig.\ref{fig:forward_reverse_population_2000}.
This plot shows that the populations of $\ket{\uparrow}$ and $\ket{\downarrow}$ are controllable by using the longitudinal magnetic field.
Fig.\ref{fig:forward_reverse_population_2000} (b) shows the population after the reverse part against $h_{z}$. 
This illustrates that we can choose any populations of $\ket{+}$ and $\ket{-}$ by changing the value of $h_z$.
Consequently, it is evident that the rotation of the X-axis with any desired rotation angle is feasible within our framework.
It is important to note that, for small $T$, the adiabaticity is not guaranteed for both the forward and reverse parts.
We discuss how the violation of the adiabaticity affects the dynamics during the implementation of the X-rotation gate
in Appendix \ref{sec:detail_numerical_simulation}.
On the other hand, as $T$ approaches $\infty$ in the asymptotic limit, the analytical solution can be derived using perturbation theory.
we discuss the result in detail in the Appendix \ref{sec:perturbative_theory}.

\subsection{Controlled-not gate}

\begin{figure*}[t]
    \centering
    \includegraphics[width=0.75\linewidth]{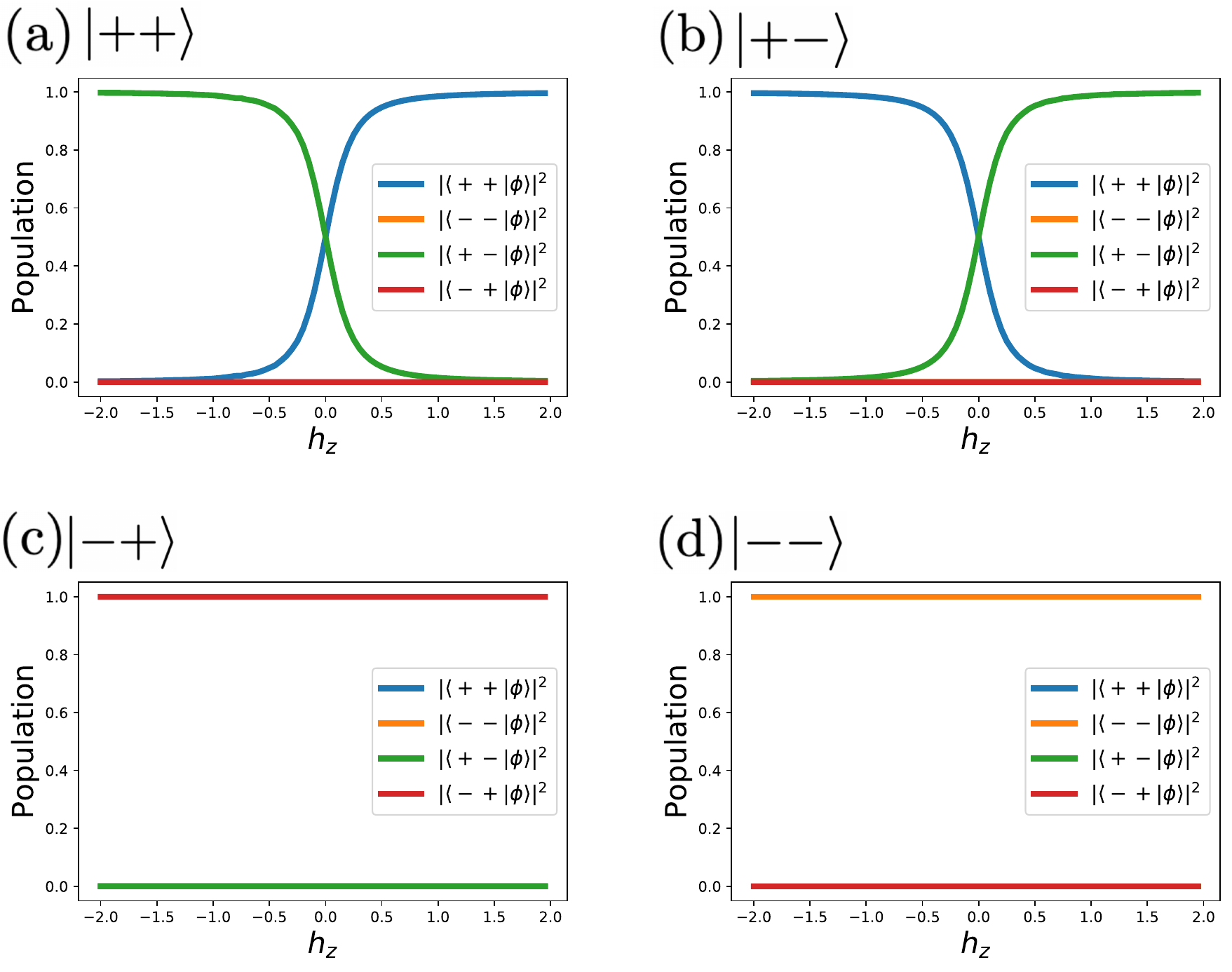}
    \caption{
    Numerical results of the controlled-not gate with our method.
    We plot the population of $\ket{++}$, $\ket{+-}$, $\ket{-+}$, and $\ket{--}$ after performing our controlled-not gate. 
    We consider the first (second) qubit as the control (target) qubit. 
    From (a) and (b), if the first qubit is $\ket{+}$, the second qubit is rotated along X axis, and we can tune the rotation angle by changing the strength of the longitudinal magnetic field $h_z$.
    On the other hand, from (c) and (d), if the first qubit is $\ket{-}$, the second qubit is unchanged.
    These results illustrate that we can implement the controlled-not gate with our method.
    }\label{fig:cnot_numerical_result_main_text}
\end{figure*}

In this subsection, we show the results of the numerical simulations of the controlled-not gate with our method.
We set the drive Hamiltonian $H_{D}$, the problem Hamiltonian $H_{P}$, and the return problem Hamiltonian $\tilde{H}_{P}$ to be those described in
(\ref{eq:c-not_drive_ham}), (\ref{eq:c-not_prob_ham}), and (\ref{eq:c-not_return_prob_ham}), respectively.
Let us assume that the initial state is one of the eigenstates of the driver Hamiltonian.
The parameters are chosen as $a=0.3$, $b=0.5$, and $T=20000$.
FIG.\ref{fig:cnot_numerical_result_main_text} illustrates the population of each eigenstate of the drive Hamiltonian, such as $\ket{++}$, $\ket{+-}$, $\ket{-+}$, and $\ket{--}$, corresponding to each initial state.
The first qubit is for control while the second qubit is the target when we perform the controlled-not gate.
We can see that if the control qubit is $\ket{+}$, the target qubit undergoes a rotating operation for an arbitrary axis. 
On the other hand, if the control qubit is $\ket{-}$, the target qubit remains unchanged.
Hence, we confirm that the controlled-not operation is successfully implemented with our method.
On the other hand, when we decrease the annealing time $T$, the adiabatic condition will be violated, and the controlled-not gate cannot be accurately implemented as shown in the Appendix.\ref{sec:detail_numerical_simulation}.

\section{Experimental demonstration}\label{sec:experimental_demonstration}
In this section, we show the results to demonstrate our method by using the D-wave device. However, due to a short coherence time, we could not observe the information of the phase in these demonstrations.  We observe the population change in the experiments, which is consistent with our theoretical prediction.
Also, since we can perform the measurement only when the transverse field is absent with the D-wave device, we perform the experiment only for the forward part of our method.
In the D-Wave device, the Hamiltonian is expressed as
\begin{align}
    H(s)=&-\frac{A(s)}{2}\biggl(\sum_{j}\hat{\sigma}_{j}^{(x)}\biggr)\notag\\
    &+\frac{B(s)}{2}\biggl(\sum_{j}g(t)h_{j}\hat{\sigma}_{j}^{(z)}+\sum_{i>j}J_{i,j}\hat{\sigma}_{i}^{(z)}\hat{\sigma}_{j}^{(z)}\biggr).\label{eq:d-wave_ham}
\end{align}
Here, we set
$A(s=1)=B(s=0)=0$.
The strength of the longitudinal magnetic field $g(t)$ is tunable.
Additionally, we can choose the values of
the longitudinal magnetic field $\{h_{j}\}_{j}$ and the coupling constant $\{J_{i,j}\}_{i>j}$.
Unfortunately, we cannot individually control the longitudinal magnetic field scheduling at each qubit with the current D-wave device. 
Thus, in this section, we replace uniform longitudinal magnetic field $H_{C}=\sum_{j}\hat{\sigma}_{j}^{{z}}$ in the Section \ref{sec:realization_of_gate_op} for the inhomogeneous longitudinal magnetic field $\sum_{j}h_{j}\hat{\sigma}_{j}^{(z)}$.
Moreover, we can prepare only the all plus state $\ket{++\cdots+}$ as an initial state in the D-Wave device.
It should be noted that we use the D-Wave Advantage system 6.3 for our demonstration.

\subsection{Rotation about the X-axis}

\begin{figure}
    \centering
    \includegraphics[width=1.0\linewidth]{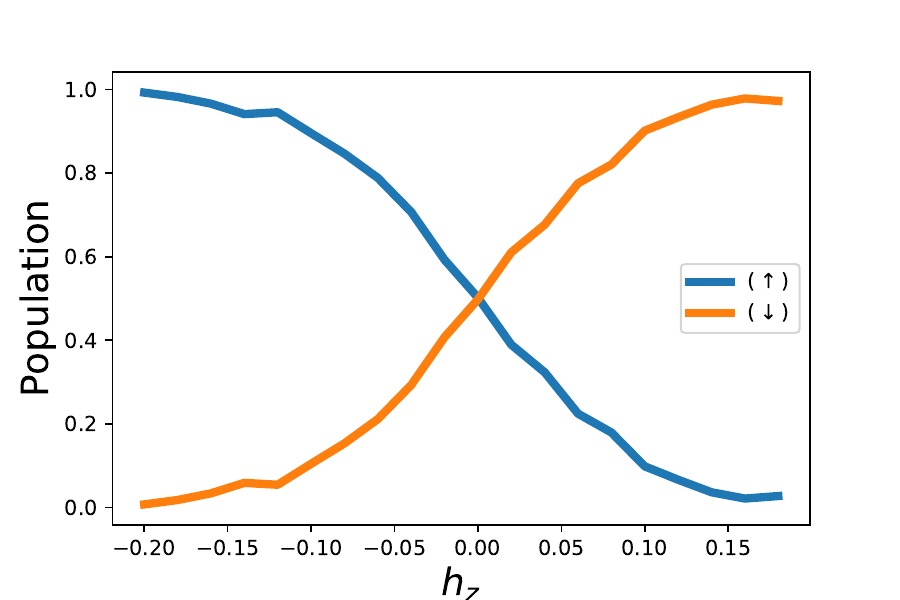}
    \caption{
    The population of each state against the amplitude parameter of the catalytic term $h_{z}$ when we perform the forward part of our X-rotation gate with the D-wave device.
    The observed behavior is consistent with our numerical results in Fig.\ref{fig:forward_reverse_population_2000} and the analytical results in Fig.\ref{fig:perturbation_plot_x_rotation_1qubit}.
    }
    \label{fig:sqr_1qubit_population_hd}
\end{figure}

We experimentally investigate the implementation of the X-rotation gate with our method using the D-wave device.
For this purpose, we need to translate
the theoretical framework of Subsection \ref{sec:realize_x_rotation} into practical experimentation. 
To realize the rotation about the X-axis using a single qubit, we set the longitudinal magnetic field as
$g(0)=g(T)=0$, $g(T/2)=h_z$, and $h_{1}=1$. 
The initial state is set to be
$\ket{\phi}=\ket{+}$.
In addition, we set the annealing time as
$T=200$, and the measurement number as $N=2000$.
Thus, we obtain the plot of the population of $\ket{\uparrow}$ and $\ket{\downarrow}$ against the strength of longitudinal magnetic field $h_{d}$ in the Fig.\ref{fig:sqr_1qubit_population_hd}.
These experimental results are consistent with our theoretical predictions about the X-rotation gate shown in Fig.\ref{fig:forward_reverse_population_2000}.

\subsection{Controlled-not gate using the D-wave device}\label{sec:c_not_d-wave}

In this subsection, we experimentally 
We experimentally investigate the implementation
of the X-rotation gate with our method using the D-wave device.
The parameters of the problem Hamiltonian are set to $a=0.3$ as defined in the equation (\ref{eq:c-not_prob_ham}).
Also, the annealing time and the number of measurements are set to be $T=200$ and $N=2000$, respectively.
In the equation (\ref{eq:d-wave_ham}), we set the strength of the longitudinal magnetic fields as $h_{1}=1$, $h_{2}=0.3$, the coupling constant as $J_{12}=0.3$.
Also, we set the scheduling of the longitudinal magnetic fields as $g(0)=0$, $g(T/2)=h_{z}+1$, and $g(T)=1$.
Due to the constraints of the D-wave device, 
we can prepare only $\ket{++}$ as the initial state.
Furthermore, we cannot apply inhomogeneous transverse magnetic fields with the D-wave device, and so we adopt the homogeneous transverse magnetic field as
\begin{align}
    H_{D}=-\sum_{j=1,2}\hat{\sigma}_{j}^{(x)}.
\end{align}

\begin{figure}
    \centering
    \includegraphics[width=1.0\linewidth]{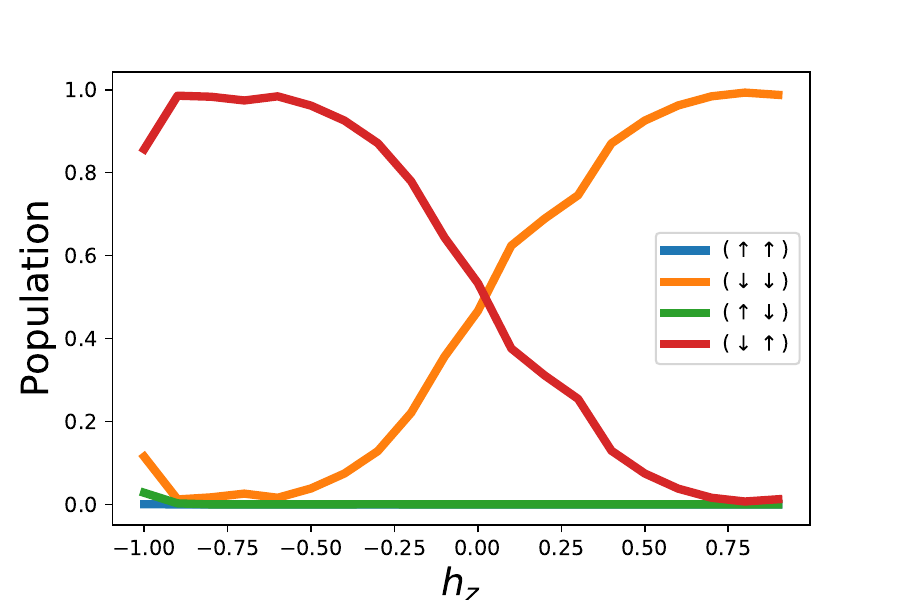}
    \caption{The population of each state against the amplitude parameter of the catalytic term $h_z$ when we perform the forward part of the controlled-not gate with the D-wave device.
    The behavior of this plot corresponds to the analytical solution as shown in Fig.\ref{fig:population_hz_a03}.
    }
    \label{fig:population_each_state1}
\end{figure}

Notably, the behavior observed in Fig \ref{fig:population_each_state1} coincides with that
of the numerical simulation depicted in Fig.\ref{fig:cnot_numerical_result_main_text}(a), demonstrating the consistency and effectiveness of our approach.

\section{Construction of the gate-based quantum computer using D-Wave device}\label{sec:construction_using_d-wave}

In this section, we construct the gate-based quantum computation using our scheme.
Initially, we consider the idling state using the transverse field.
To neglect the relative phase, we consider only the case where time is an integer multiple of $2\pi$ divided by the energy gap $k$ as follows.
\begin{align}
    t=\frac{2\pi}{k}n\ \ n\in \mathbb{Z}.
\end{align}
Subsequently, to operate the controlled-not gate and the rotation about the X-axis using previous sections \ref{sec:realize_x_rotation} and \ref{sec:propose_c-not}, we execute the gate operation through the time evolution between the transverse magnetic field and the Ising model as (\ref{eq:x-axis}) and (\ref{eq:c-not_prob_ham}).
When realizing the rotation about the Z-axis the energy difference is controlled by changing the intensity of the transverse magnetic field. 
The phase difference, as discussed in Section \ref{sec:z-axis_op}, is utilized to control this energy difference. 
Finally, we implement repeatedly these gate operations to realize the desired quantum algorithm.

\section{Conclusion}\label{sec:conclusion}

In this paper, we propose an innovative approach to implement the controlled-not gate and single-qubit rotation by using a quantum annealer with the transverse-field Ising Hamiltonian. 
Unlike the previous approach to perform the gate operations by either applying an oscillating field or accumulating a relative phase, we utilize the degenerate ground states to perform the X-rotation gate and controlled-not gate.
We demonstrate the effectiveness of our method by numerical simulations and experimental results.
Importantly, we do not use microwave pulses to implement gate operations.
This feature enhances scalability, as evidenced by the large-scale devices developed by D-Wave Inc.
Employing only the transverse-field Ising model, our approach holds substantial promise for developing large-scale gate-based quantum computers.

\begin{acknowledgments}
This paper is partly
based on the results obtained from a project, JPNP16007,
commissioned by the New Energy and Industrial Technology Development Organization (NEDO), Japan.
\end{acknowledgments}

\nocite{*}

\bibliography{apssamp}

\providecommand{\noopsort}[1]{}\providecommand{\singleletter}[1]{#1}%
\begin{thebibliography}{69}%
\makeatletter
\providecommand \@ifxundefined [1]{%
 \@ifx{#1\undefined}
}%
\providecommand \@ifnum [1]{%
 \ifnum #1\expandafter \@firstoftwo
 \else \expandafter \@secondoftwo
 \fi
}%
\providecommand \@ifx [1]{%
 \ifx #1\expandafter \@firstoftwo
 \else \expandafter \@secondoftwo
 \fi
}%
\providecommand \natexlab [1]{#1}%
\providecommand \enquote  [1]{``#1''}%
\providecommand \bibnamefont  [1]{#1}%
\providecommand \bibfnamefont [1]{#1}%
\providecommand \citenamefont [1]{#1}%
\providecommand \href@noop [0]{\@secondoftwo}%
\providecommand \href [0]{\begingroup \@sanitize@url \@href}%
\providecommand \@href[1]{\@@startlink{#1}\@@href}%
\providecommand \@@href[1]{\endgroup#1\@@endlink}%
\providecommand \@sanitize@url [0]{\catcode `\\12\catcode `\$12\catcode
  `\&12\catcode `\#12\catcode `\^12\catcode `\_12\catcode `\%12\relax}%
\providecommand \@@startlink[1]{}%
\providecommand \@@endlink[0]{}%
\providecommand \url  [0]{\begingroup\@sanitize@url \@url }%
\providecommand \@url [1]{\endgroup\@href {#1}{\urlprefix }}%
\providecommand \urlprefix  [0]{URL }%
\providecommand \Eprint [0]{\href }%
\providecommand \doibase [0]{https://doi.org/}%
\providecommand \selectlanguage [0]{\@gobble}%
\providecommand \bibinfo  [0]{\@secondoftwo}%
\providecommand \bibfield  [0]{\@secondoftwo}%
\providecommand \translation [1]{[#1]}%
\providecommand \BibitemOpen [0]{}%
\providecommand \bibitemStop [0]{}%
\providecommand \bibitemNoStop [0]{.\EOS\space}%
\providecommand \EOS [0]{\spacefactor3000\relax}%
\providecommand \BibitemShut  [1]{\csname bibitem#1\endcsname}%
\let\auto@bib@innerbib\@empty
\bibitem [{\citenamefont {Kadowaki}\ and\ \citenamefont
  {Nishimori}(1998)}]{kadowaki1998quantum}%
  \BibitemOpen
  \bibfield  {author} {\bibinfo {author} {\bibfnamefont {T.}~\bibnamefont
  {Kadowaki}}\ and\ \bibinfo {author} {\bibfnamefont {H.}~\bibnamefont
  {Nishimori}},\ }\bibfield  {title} {\bibinfo {title} {Quantum annealing in
  the transverse ising model},\ }\href@noop {} {\bibfield  {journal} {\bibinfo
  {journal} {Physical Review E}\ }\textbf {\bibinfo {volume} {58}},\ \bibinfo
  {pages} {5355} (\bibinfo {year} {1998})}\BibitemShut {NoStop}%
\bibitem [{\citenamefont {Farhi}\ \emph {et~al.}(2000)\citenamefont {Farhi},
  \citenamefont {Goldstone}, \citenamefont {Gutmann},\ and\ \citenamefont
  {Sipser}}]{farhi2000quantum}%
  \BibitemOpen
  \bibfield  {author} {\bibinfo {author} {\bibfnamefont {E.}~\bibnamefont
  {Farhi}}, \bibinfo {author} {\bibfnamefont {J.}~\bibnamefont {Goldstone}},
  \bibinfo {author} {\bibfnamefont {S.}~\bibnamefont {Gutmann}},\ and\ \bibinfo
  {author} {\bibfnamefont {M.}~\bibnamefont {Sipser}},\ }\bibfield  {title}
  {\bibinfo {title} {Quantum computation by adiabatic evolution},\ }\href@noop
  {} {\bibfield  {journal} {\bibinfo  {journal} {arXiv preprint
  quant-ph/0001106}\ } (\bibinfo {year} {2000})}\BibitemShut {NoStop}%
\bibitem [{\citenamefont {Farhi}\ \emph {et~al.}(2001)\citenamefont {Farhi},
  \citenamefont {Goldstone}, \citenamefont {Gutmann}, \citenamefont {Lapan},
  \citenamefont {Lundgren},\ and\ \citenamefont {Preda}}]{farhi2001quantum}%
  \BibitemOpen
  \bibfield  {author} {\bibinfo {author} {\bibfnamefont {E.}~\bibnamefont
  {Farhi}}, \bibinfo {author} {\bibfnamefont {J.}~\bibnamefont {Goldstone}},
  \bibinfo {author} {\bibfnamefont {S.}~\bibnamefont {Gutmann}}, \bibinfo
  {author} {\bibfnamefont {J.}~\bibnamefont {Lapan}}, \bibinfo {author}
  {\bibfnamefont {A.}~\bibnamefont {Lundgren}},\ and\ \bibinfo {author}
  {\bibfnamefont {D.}~\bibnamefont {Preda}},\ }\bibfield  {title} {\bibinfo
  {title} {A quantum adiabatic evolution algorithm applied to random instances
  of an np-complete problem},\ }\href@noop {} {\bibfield  {journal} {\bibinfo
  {journal} {Science}\ }\textbf {\bibinfo {volume} {292}},\ \bibinfo {pages}
  {472} (\bibinfo {year} {2001})}\BibitemShut {NoStop}%
\bibitem [{\citenamefont {Barenco}\ \emph {et~al.}(1995)\citenamefont
  {Barenco}, \citenamefont {Bennett}, \citenamefont {Cleve}, \citenamefont
  {DiVincenzo}, \citenamefont {Margolus}, \citenamefont {Shor}, \citenamefont
  {Sleator}, \citenamefont {Smolin},\ and\ \citenamefont
  {Weinfurter}}]{barenco1995elementary}%
  \BibitemOpen
  \bibfield  {author} {\bibinfo {author} {\bibfnamefont {A.}~\bibnamefont
  {Barenco}}, \bibinfo {author} {\bibfnamefont {C.~H.}\ \bibnamefont
  {Bennett}}, \bibinfo {author} {\bibfnamefont {R.}~\bibnamefont {Cleve}},
  \bibinfo {author} {\bibfnamefont {D.~P.}\ \bibnamefont {DiVincenzo}},
  \bibinfo {author} {\bibfnamefont {N.}~\bibnamefont {Margolus}}, \bibinfo
  {author} {\bibfnamefont {P.}~\bibnamefont {Shor}}, \bibinfo {author}
  {\bibfnamefont {T.}~\bibnamefont {Sleator}}, \bibinfo {author} {\bibfnamefont
  {J.~A.}\ \bibnamefont {Smolin}},\ and\ \bibinfo {author} {\bibfnamefont
  {H.}~\bibnamefont {Weinfurter}},\ }\bibfield  {title} {\bibinfo {title}
  {Elementary gates for quantum computation},\ }\href@noop {} {\bibfield
  {journal} {\bibinfo  {journal} {Physical review A}\ }\textbf {\bibinfo
  {volume} {52}},\ \bibinfo {pages} {3457} (\bibinfo {year}
  {1995})}\BibitemShut {NoStop}%
\bibitem [{\citenamefont {Boykin}\ \emph {et~al.}(2000)\citenamefont {Boykin},
  \citenamefont {Mor}, \citenamefont {Pulver}, \citenamefont {Roychowdhury},\
  and\ \citenamefont {Vatan}}]{boykin2000new}%
  \BibitemOpen
  \bibfield  {author} {\bibinfo {author} {\bibfnamefont {P.~O.}\ \bibnamefont
  {Boykin}}, \bibinfo {author} {\bibfnamefont {T.}~\bibnamefont {Mor}},
  \bibinfo {author} {\bibfnamefont {M.}~\bibnamefont {Pulver}}, \bibinfo
  {author} {\bibfnamefont {V.}~\bibnamefont {Roychowdhury}},\ and\ \bibinfo
  {author} {\bibfnamefont {F.}~\bibnamefont {Vatan}},\ }\bibfield  {title}
  {\bibinfo {title} {A new universal and fault-tolerant quantum basis},\
  }\href@noop {} {\bibfield  {journal} {\bibinfo  {journal} {Information
  Processing Letters}\ }\textbf {\bibinfo {volume} {75}},\ \bibinfo {pages}
  {101} (\bibinfo {year} {2000})}\BibitemShut {NoStop}%
\bibitem [{\citenamefont {Kitaev}(1995)}]{kitaev1995quantum}%
  \BibitemOpen
  \bibfield  {author} {\bibinfo {author} {\bibfnamefont {A.~Y.}\ \bibnamefont
  {Kitaev}},\ }\bibfield  {title} {\bibinfo {title} {Quantum measurements and
  the abelian stabilizer problem},\ }\href@noop {} {\bibfield  {journal}
  {\bibinfo  {journal} {arXiv preprint quant-ph/9511026}\ } (\bibinfo {year}
  {1995})}\BibitemShut {NoStop}%
\bibitem [{\citenamefont {Shi}(2002)}]{shi2002both}%
  \BibitemOpen
  \bibfield  {author} {\bibinfo {author} {\bibfnamefont {Y.}~\bibnamefont
  {Shi}},\ }\bibfield  {title} {\bibinfo {title} {Both toffoli and
  controlled-not need little help to do universal quantum computation},\
  }\href@noop {} {\bibfield  {journal} {\bibinfo  {journal} {arXiv preprint
  quant-ph/0205115}\ } (\bibinfo {year} {2002})}\BibitemShut {NoStop}%
\bibitem [{\citenamefont {Shor}(1994)}]{shor1994algorithms}%
  \BibitemOpen
  \bibfield  {author} {\bibinfo {author} {\bibfnamefont {P.~W.}\ \bibnamefont
  {Shor}},\ }\bibfield  {title} {\bibinfo {title} {Algorithms for quantum
  computation: discrete logarithms and factoring},\ }in\ \href@noop {} {\emph
  {\bibinfo {booktitle} {Proceedings 35th annual symposium on foundations of
  computer science}}}\ (\bibinfo {organization} {Ieee},\ \bibinfo {year}
  {1994})\ pp.\ \bibinfo {pages} {124--134}\BibitemShut {NoStop}%
\bibitem [{\citenamefont {Shor}(1999)}]{shor1999polynomial}%
  \BibitemOpen
  \bibfield  {author} {\bibinfo {author} {\bibfnamefont {P.~W.}\ \bibnamefont
  {Shor}},\ }\bibfield  {title} {\bibinfo {title} {Polynomial-time algorithms
  for prime factorization and discrete logarithms on a quantum computer},\
  }\href@noop {} {\bibfield  {journal} {\bibinfo  {journal} {SIAM review}\
  }\textbf {\bibinfo {volume} {41}},\ \bibinfo {pages} {303} (\bibinfo {year}
  {1999})}\BibitemShut {NoStop}%
\bibitem [{\citenamefont {Harrow}\ \emph {et~al.}(2009)\citenamefont {Harrow},
  \citenamefont {Hassidim},\ and\ \citenamefont {Lloyd}}]{harrow2009quantum}%
  \BibitemOpen
  \bibfield  {author} {\bibinfo {author} {\bibfnamefont {A.~W.}\ \bibnamefont
  {Harrow}}, \bibinfo {author} {\bibfnamefont {A.}~\bibnamefont {Hassidim}},\
  and\ \bibinfo {author} {\bibfnamefont {S.}~\bibnamefont {Lloyd}},\ }\bibfield
   {title} {\bibinfo {title} {Quantum algorithm for linear systems of
  equations},\ }\href@noop {} {\bibfield  {journal} {\bibinfo  {journal}
  {Physical review letters}\ }\textbf {\bibinfo {volume} {103}},\ \bibinfo
  {pages} {150502} (\bibinfo {year} {2009})}\BibitemShut {NoStop}%
\bibitem [{\citenamefont {Nielsen}\ and\ \citenamefont
  {Chuang}(2010)}]{nielsen2010quantum}%
  \BibitemOpen
  \bibfield  {author} {\bibinfo {author} {\bibfnamefont {M.~A.}\ \bibnamefont
  {Nielsen}}\ and\ \bibinfo {author} {\bibfnamefont {I.~L.}\ \bibnamefont
  {Chuang}},\ }\href@noop {} {\emph {\bibinfo {title} {Quantum computation and
  quantum information}}}\ (\bibinfo  {publisher} {Cambridge university press},\
  \bibinfo {year} {2010})\BibitemShut {NoStop}%
\bibitem [{\citenamefont {Grover}(1996)}]{grover1996fast}%
  \BibitemOpen
  \bibfield  {author} {\bibinfo {author} {\bibfnamefont {L.~K.}\ \bibnamefont
  {Grover}},\ }\bibfield  {title} {\bibinfo {title} {A fast quantum mechanical
  algorithm for database search},\ }in\ \href@noop {} {\emph {\bibinfo
  {booktitle} {Proceedings of the twenty-eighth annual ACM symposium on Theory
  of computing}}}\ (\bibinfo {year} {1996})\ pp.\ \bibinfo {pages}
  {212--219}\BibitemShut {NoStop}%
\bibitem [{\citenamefont {Kitaev}(1997)}]{kitaev1997quantum}%
  \BibitemOpen
  \bibfield  {author} {\bibinfo {author} {\bibfnamefont {A.~Y.}\ \bibnamefont
  {Kitaev}},\ }\bibfield  {title} {\bibinfo {title} {Quantum computations:
  algorithms and error correction},\ }\href@noop {} {\bibfield  {journal}
  {\bibinfo  {journal} {Russian Mathematical Surveys}\ }\textbf {\bibinfo
  {volume} {52}},\ \bibinfo {pages} {1191} (\bibinfo {year}
  {1997})}\BibitemShut {NoStop}%
\bibitem [{\citenamefont {Boixo}\ \emph {et~al.}(2014)\citenamefont {Boixo},
  \citenamefont {R{\o}nnow}, \citenamefont {Isakov}, \citenamefont {Wang},
  \citenamefont {Wecker}, \citenamefont {Lidar}, \citenamefont {Martinis},\
  and\ \citenamefont {Troyer}}]{boixo2014evidence}%
  \BibitemOpen
  \bibfield  {author} {\bibinfo {author} {\bibfnamefont {S.}~\bibnamefont
  {Boixo}}, \bibinfo {author} {\bibfnamefont {T.~F.}\ \bibnamefont
  {R{\o}nnow}}, \bibinfo {author} {\bibfnamefont {S.~V.}\ \bibnamefont
  {Isakov}}, \bibinfo {author} {\bibfnamefont {Z.}~\bibnamefont {Wang}},
  \bibinfo {author} {\bibfnamefont {D.}~\bibnamefont {Wecker}}, \bibinfo
  {author} {\bibfnamefont {D.~A.}\ \bibnamefont {Lidar}}, \bibinfo {author}
  {\bibfnamefont {J.~M.}\ \bibnamefont {Martinis}},\ and\ \bibinfo {author}
  {\bibfnamefont {M.}~\bibnamefont {Troyer}},\ }\bibfield  {title} {\bibinfo
  {title} {Evidence for quantum annealing with more than one hundred qubits},\
  }\href@noop {} {\bibfield  {journal} {\bibinfo  {journal} {Nature physics}\
  }\textbf {\bibinfo {volume} {10}},\ \bibinfo {pages} {218} (\bibinfo {year}
  {2014})}\BibitemShut {NoStop}%
\bibitem [{\citenamefont {Johnson}\ \emph {et~al.}(2011)\citenamefont
  {Johnson}, \citenamefont {Amin}, \citenamefont {Gildert}, \citenamefont
  {Lanting}, \citenamefont {Hamze}, \citenamefont {Dickson}, \citenamefont
  {Harris}, \citenamefont {Berkley}, \citenamefont {Johansson}, \citenamefont
  {Bunyk} \emph {et~al.}}]{johnson2011quantum}%
  \BibitemOpen
  \bibfield  {author} {\bibinfo {author} {\bibfnamefont {M.~W.}\ \bibnamefont
  {Johnson}}, \bibinfo {author} {\bibfnamefont {M.~H.}\ \bibnamefont {Amin}},
  \bibinfo {author} {\bibfnamefont {S.}~\bibnamefont {Gildert}}, \bibinfo
  {author} {\bibfnamefont {T.}~\bibnamefont {Lanting}}, \bibinfo {author}
  {\bibfnamefont {F.}~\bibnamefont {Hamze}}, \bibinfo {author} {\bibfnamefont
  {N.}~\bibnamefont {Dickson}}, \bibinfo {author} {\bibfnamefont
  {R.}~\bibnamefont {Harris}}, \bibinfo {author} {\bibfnamefont {A.~J.}\
  \bibnamefont {Berkley}}, \bibinfo {author} {\bibfnamefont {J.}~\bibnamefont
  {Johansson}}, \bibinfo {author} {\bibfnamefont {P.}~\bibnamefont {Bunyk}},
  \emph {et~al.},\ }\bibfield  {title} {\bibinfo {title} {Quantum annealing
  with manufactured spins},\ }\href@noop {} {\bibfield  {journal} {\bibinfo
  {journal} {Nature}\ }\textbf {\bibinfo {volume} {473}},\ \bibinfo {pages}
  {194} (\bibinfo {year} {2011})}\BibitemShut {NoStop}%
\bibitem [{\citenamefont {Adachi}\ and\ \citenamefont
  {Henderson}(2015)}]{adachi2015application}%
  \BibitemOpen
  \bibfield  {author} {\bibinfo {author} {\bibfnamefont {S.~H.}\ \bibnamefont
  {Adachi}}\ and\ \bibinfo {author} {\bibfnamefont {M.~P.}\ \bibnamefont
  {Henderson}},\ }\bibfield  {title} {\bibinfo {title} {Application of quantum
  annealing to training of deep neural networks},\ }\href@noop {} {\bibfield
  {journal} {\bibinfo  {journal} {arXiv preprint arXiv:1510.06356}\ } (\bibinfo
  {year} {2015})}\BibitemShut {NoStop}%
\bibitem [{\citenamefont {Sasdelli}\ and\ \citenamefont
  {Chin}(2021)}]{sasdelli2021quantum}%
  \BibitemOpen
  \bibfield  {author} {\bibinfo {author} {\bibfnamefont {M.}~\bibnamefont
  {Sasdelli}}\ and\ \bibinfo {author} {\bibfnamefont {T.-J.}\ \bibnamefont
  {Chin}},\ }\bibfield  {title} {\bibinfo {title} {Quantum annealing
  formulation for binary neural networks},\ }in\ \href@noop {} {\emph {\bibinfo
  {booktitle} {2021 Digital Image Computing: Techniques and Applications
  (DICTA)}}}\ (\bibinfo {organization} {IEEE},\ \bibinfo {year} {2021})\ pp.\
  \bibinfo {pages} {1--10}\BibitemShut {NoStop}%
\bibitem [{\citenamefont {Date}\ and\ \citenamefont
  {Potok}(2021)}]{date2021adiabatic}%
  \BibitemOpen
  \bibfield  {author} {\bibinfo {author} {\bibfnamefont {P.}~\bibnamefont
  {Date}}\ and\ \bibinfo {author} {\bibfnamefont {T.}~\bibnamefont {Potok}},\
  }\bibfield  {title} {\bibinfo {title} {Adiabatic quantum linear regression},\
  }\href@noop {} {\bibfield  {journal} {\bibinfo  {journal} {Scientific
  reports}\ }\textbf {\bibinfo {volume} {11}},\ \bibinfo {pages} {21905}
  (\bibinfo {year} {2021})}\BibitemShut {NoStop}%
\bibitem [{\citenamefont {Wilson}\ \emph {et~al.}(2021)\citenamefont {Wilson},
  \citenamefont {Vandal}, \citenamefont {Hogg},\ and\ \citenamefont
  {Rieffel}}]{wilson2021quantum}%
  \BibitemOpen
  \bibfield  {author} {\bibinfo {author} {\bibfnamefont {M.}~\bibnamefont
  {Wilson}}, \bibinfo {author} {\bibfnamefont {T.}~\bibnamefont {Vandal}},
  \bibinfo {author} {\bibfnamefont {T.}~\bibnamefont {Hogg}},\ and\ \bibinfo
  {author} {\bibfnamefont {E.~G.}\ \bibnamefont {Rieffel}},\ }\bibfield
  {title} {\bibinfo {title} {Quantum-assisted associative adversarial network:
  Applying quantum annealing in deep learning},\ }\href@noop {} {\bibfield
  {journal} {\bibinfo  {journal} {Quantum Machine Intelligence}\ }\textbf
  {\bibinfo {volume} {3}},\ \bibinfo {pages} {1} (\bibinfo {year}
  {2021})}\BibitemShut {NoStop}%
\bibitem [{\citenamefont {Kumar}\ \emph {et~al.}(2018)\citenamefont {Kumar},
  \citenamefont {Bass}, \citenamefont {Tomlin},\ and\ \citenamefont
  {Dulny}}]{kumar2018quantum}%
  \BibitemOpen
  \bibfield  {author} {\bibinfo {author} {\bibfnamefont {V.}~\bibnamefont
  {Kumar}}, \bibinfo {author} {\bibfnamefont {G.}~\bibnamefont {Bass}},
  \bibinfo {author} {\bibfnamefont {C.}~\bibnamefont {Tomlin}},\ and\ \bibinfo
  {author} {\bibfnamefont {J.}~\bibnamefont {Dulny}},\ }\bibfield  {title}
  {\bibinfo {title} {Quantum annealing for combinatorial clustering},\
  }\href@noop {} {\bibfield  {journal} {\bibinfo  {journal} {Quantum
  Information Processing}\ }\textbf {\bibinfo {volume} {17}},\ \bibinfo {pages}
  {1} (\bibinfo {year} {2018})}\BibitemShut {NoStop}%
\bibitem [{\citenamefont {Kurihara}\ \emph {et~al.}(2014)\citenamefont
  {Kurihara}, \citenamefont {Tanaka},\ and\ \citenamefont
  {Miyashita}}]{kurihara2014quantum}%
  \BibitemOpen
  \bibfield  {author} {\bibinfo {author} {\bibfnamefont {K.}~\bibnamefont
  {Kurihara}}, \bibinfo {author} {\bibfnamefont {S.}~\bibnamefont {Tanaka}},\
  and\ \bibinfo {author} {\bibfnamefont {S.}~\bibnamefont {Miyashita}},\
  }\bibfield  {title} {\bibinfo {title} {Quantum annealing for clustering},\
  }\href@noop {} {\bibfield  {journal} {\bibinfo  {journal} {arXiv preprint
  arXiv:1408.2035}\ } (\bibinfo {year} {2014})}\BibitemShut {NoStop}%
\bibitem [{\citenamefont {Willsch}\ \emph {et~al.}(2020)\citenamefont
  {Willsch}, \citenamefont {Willsch}, \citenamefont {De~Raedt},\ and\
  \citenamefont {Michielsen}}]{willsch2020support}%
  \BibitemOpen
  \bibfield  {author} {\bibinfo {author} {\bibfnamefont {D.}~\bibnamefont
  {Willsch}}, \bibinfo {author} {\bibfnamefont {M.}~\bibnamefont {Willsch}},
  \bibinfo {author} {\bibfnamefont {H.}~\bibnamefont {De~Raedt}},\ and\
  \bibinfo {author} {\bibfnamefont {K.}~\bibnamefont {Michielsen}},\ }\bibfield
   {title} {\bibinfo {title} {Support vector machines on the d-wave quantum
  annealer},\ }\href@noop {} {\bibfield  {journal} {\bibinfo  {journal}
  {Computer physics communications}\ }\textbf {\bibinfo {volume} {248}},\
  \bibinfo {pages} {107006} (\bibinfo {year} {2020})}\BibitemShut {NoStop}%
\bibitem [{\citenamefont {Heidari}\ \emph {et~al.}(2023)\citenamefont
  {Heidari}, \citenamefont {Dinneen},\ and\ \citenamefont
  {Delmas}}]{heidari2023quantum}%
  \BibitemOpen
  \bibfield  {author} {\bibinfo {author} {\bibfnamefont {S.}~\bibnamefont
  {Heidari}}, \bibinfo {author} {\bibfnamefont {M.~J.}\ \bibnamefont
  {Dinneen}},\ and\ \bibinfo {author} {\bibfnamefont {P.}~\bibnamefont
  {Delmas}},\ }\bibfield  {title} {\bibinfo {title} {Quantum annealing for
  computer vision minimization problems},\ }\href@noop {} {\bibfield  {journal}
  {\bibinfo  {journal} {arXiv preprint arXiv:2312.12848}\ } (\bibinfo {year}
  {2023})}\BibitemShut {NoStop}%
\bibitem [{\citenamefont {Woun}\ and\ \citenamefont
  {Date}(2023)}]{woun2023adiabatic}%
  \BibitemOpen
  \bibfield  {author} {\bibinfo {author} {\bibfnamefont {D.~J.}\ \bibnamefont
  {Woun}}\ and\ \bibinfo {author} {\bibfnamefont {P.}~\bibnamefont {Date}},\
  }\bibfield  {title} {\bibinfo {title} {Adiabatic quantum support vector
  machines},\ }in\ \href@noop {} {\emph {\bibinfo {booktitle} {2023 IEEE
  International Conference on Quantum Computing and Engineering (QCE)}}},\
  Vol.~\bibinfo {volume} {2}\ (\bibinfo {organization} {IEEE},\ \bibinfo {year}
  {2023})\ pp.\ \bibinfo {pages} {296--297}\BibitemShut {NoStop}%
\bibitem [{\citenamefont {Neven}\ \emph {et~al.}(2009)\citenamefont {Neven},
  \citenamefont {Denchev}, \citenamefont {Rose},\ and\ \citenamefont
  {Macready}}]{neven2009training}%
  \BibitemOpen
  \bibfield  {author} {\bibinfo {author} {\bibfnamefont {H.}~\bibnamefont
  {Neven}}, \bibinfo {author} {\bibfnamefont {V.~S.}\ \bibnamefont {Denchev}},
  \bibinfo {author} {\bibfnamefont {G.}~\bibnamefont {Rose}},\ and\ \bibinfo
  {author} {\bibfnamefont {W.~G.}\ \bibnamefont {Macready}},\ }\bibfield
  {title} {\bibinfo {title} {Training a large scale classifier with the quantum
  adiabatic algorithm},\ }\href@noop {} {\bibfield  {journal} {\bibinfo
  {journal} {arXiv preprint arXiv:0912.0779}\ } (\bibinfo {year}
  {2009})}\BibitemShut {NoStop}%
\bibitem [{\citenamefont {Bosch}\ \emph {et~al.}(2023)\citenamefont {Bosch},
  \citenamefont {Kiani}, \citenamefont {Yang}, \citenamefont {Lupascu},\ and\
  \citenamefont {Lloyd}}]{bosch2023neural}%
  \BibitemOpen
  \bibfield  {author} {\bibinfo {author} {\bibfnamefont {S.}~\bibnamefont
  {Bosch}}, \bibinfo {author} {\bibfnamefont {B.}~\bibnamefont {Kiani}},
  \bibinfo {author} {\bibfnamefont {R.}~\bibnamefont {Yang}}, \bibinfo {author}
  {\bibfnamefont {A.}~\bibnamefont {Lupascu}},\ and\ \bibinfo {author}
  {\bibfnamefont {S.}~\bibnamefont {Lloyd}},\ }\bibfield  {title} {\bibinfo
  {title} {Neural networks for programming quantum annealers},\ }\href@noop {}
  {\bibfield  {journal} {\bibinfo  {journal} {arXiv preprint arXiv:2308.06807}\
  } (\bibinfo {year} {2023})}\BibitemShut {NoStop}%
\bibitem [{\citenamefont {Pudenz}\ and\ \citenamefont
  {Lidar}(2013)}]{pudenz2013quantum}%
  \BibitemOpen
  \bibfield  {author} {\bibinfo {author} {\bibfnamefont {K.~L.}\ \bibnamefont
  {Pudenz}}\ and\ \bibinfo {author} {\bibfnamefont {D.~A.}\ \bibnamefont
  {Lidar}},\ }\bibfield  {title} {\bibinfo {title} {Quantum adiabatic machine
  learning},\ }\href@noop {} {\bibfield  {journal} {\bibinfo  {journal}
  {Quantum information processing}\ }\textbf {\bibinfo {volume} {12}},\
  \bibinfo {pages} {2027} (\bibinfo {year} {2013})}\BibitemShut {NoStop}%
\bibitem [{\citenamefont {Zhou}\ \emph {et~al.}(2021)\citenamefont {Zhou},
  \citenamefont {Green}, \citenamefont {Dahl},\ and\ \citenamefont
  {Chamon}}]{zhou2021experimental}%
  \BibitemOpen
  \bibfield  {author} {\bibinfo {author} {\bibfnamefont {S.}~\bibnamefont
  {Zhou}}, \bibinfo {author} {\bibfnamefont {D.}~\bibnamefont {Green}},
  \bibinfo {author} {\bibfnamefont {E.~D.}\ \bibnamefont {Dahl}},\ and\
  \bibinfo {author} {\bibfnamefont {C.}~\bibnamefont {Chamon}},\ }\bibfield
  {title} {\bibinfo {title} {Experimental realization of classical z 2 spin
  liquids in a programmable quantum device},\ }\href@noop {} {\bibfield
  {journal} {\bibinfo  {journal} {Physical Review B}\ }\textbf {\bibinfo
  {volume} {104}},\ \bibinfo {pages} {L081107} (\bibinfo {year}
  {2021})}\BibitemShut {NoStop}%
\bibitem [{\citenamefont {Kairys}\ \emph {et~al.}(2020)\citenamefont {Kairys},
  \citenamefont {King}, \citenamefont {Ozfidan}, \citenamefont {Boothby},
  \citenamefont {Raymond}, \citenamefont {Banerjee},\ and\ \citenamefont
  {Humble}}]{kairys2020simulating}%
  \BibitemOpen
  \bibfield  {author} {\bibinfo {author} {\bibfnamefont {P.}~\bibnamefont
  {Kairys}}, \bibinfo {author} {\bibfnamefont {A.~D.}\ \bibnamefont {King}},
  \bibinfo {author} {\bibfnamefont {I.}~\bibnamefont {Ozfidan}}, \bibinfo
  {author} {\bibfnamefont {K.}~\bibnamefont {Boothby}}, \bibinfo {author}
  {\bibfnamefont {J.}~\bibnamefont {Raymond}}, \bibinfo {author} {\bibfnamefont
  {A.}~\bibnamefont {Banerjee}},\ and\ \bibinfo {author} {\bibfnamefont
  {T.~S.}\ \bibnamefont {Humble}},\ }\bibfield  {title} {\bibinfo {title}
  {Simulating the shastry-sutherland ising model using quantum annealing},\
  }\href@noop {} {\bibfield  {journal} {\bibinfo  {journal} {Prx Quantum}\
  }\textbf {\bibinfo {volume} {1}},\ \bibinfo {pages} {020320} (\bibinfo {year}
  {2020})}\BibitemShut {NoStop}%
\bibitem [{\citenamefont {Harris}\ \emph {et~al.}(2018)\citenamefont {Harris},
  \citenamefont {Sato}, \citenamefont {Berkley}, \citenamefont {Reis},
  \citenamefont {Altomare}, \citenamefont {Amin}, \citenamefont {Boothby},
  \citenamefont {Bunyk}, \citenamefont {Deng}, \citenamefont {Enderud} \emph
  {et~al.}}]{harris2018phase}%
  \BibitemOpen
  \bibfield  {author} {\bibinfo {author} {\bibfnamefont {R.}~\bibnamefont
  {Harris}}, \bibinfo {author} {\bibfnamefont {Y.}~\bibnamefont {Sato}},
  \bibinfo {author} {\bibfnamefont {A.~J.}\ \bibnamefont {Berkley}}, \bibinfo
  {author} {\bibfnamefont {M.}~\bibnamefont {Reis}}, \bibinfo {author}
  {\bibfnamefont {F.}~\bibnamefont {Altomare}}, \bibinfo {author}
  {\bibfnamefont {M.}~\bibnamefont {Amin}}, \bibinfo {author} {\bibfnamefont
  {K.}~\bibnamefont {Boothby}}, \bibinfo {author} {\bibfnamefont
  {P.}~\bibnamefont {Bunyk}}, \bibinfo {author} {\bibfnamefont
  {C.}~\bibnamefont {Deng}}, \bibinfo {author} {\bibfnamefont {C.}~\bibnamefont
  {Enderud}}, \emph {et~al.},\ }\bibfield  {title} {\bibinfo {title} {Phase
  transitions in a programmable quantum spin glass simulator},\ }\href@noop {}
  {\bibfield  {journal} {\bibinfo  {journal} {Science}\ }\textbf {\bibinfo
  {volume} {361}},\ \bibinfo {pages} {162} (\bibinfo {year}
  {2018})}\BibitemShut {NoStop}%
\bibitem [{\citenamefont {King}\ \emph {et~al.}(2018)\citenamefont {King},
  \citenamefont {Carrasquilla}, \citenamefont {Raymond}, \citenamefont
  {Ozfidan}, \citenamefont {Andriyash}, \citenamefont {Berkley}, \citenamefont
  {Reis}, \citenamefont {Lanting}, \citenamefont {Harris}, \citenamefont
  {Altomare} \emph {et~al.}}]{king2018observation}%
  \BibitemOpen
  \bibfield  {author} {\bibinfo {author} {\bibfnamefont {A.~D.}\ \bibnamefont
  {King}}, \bibinfo {author} {\bibfnamefont {J.}~\bibnamefont {Carrasquilla}},
  \bibinfo {author} {\bibfnamefont {J.}~\bibnamefont {Raymond}}, \bibinfo
  {author} {\bibfnamefont {I.}~\bibnamefont {Ozfidan}}, \bibinfo {author}
  {\bibfnamefont {E.}~\bibnamefont {Andriyash}}, \bibinfo {author}
  {\bibfnamefont {A.}~\bibnamefont {Berkley}}, \bibinfo {author} {\bibfnamefont
  {M.}~\bibnamefont {Reis}}, \bibinfo {author} {\bibfnamefont {T.}~\bibnamefont
  {Lanting}}, \bibinfo {author} {\bibfnamefont {R.}~\bibnamefont {Harris}},
  \bibinfo {author} {\bibfnamefont {F.}~\bibnamefont {Altomare}}, \emph
  {et~al.},\ }\bibfield  {title} {\bibinfo {title} {Observation of topological
  phenomena in a programmable lattice of 1,800 qubits},\ }\href@noop {}
  {\bibfield  {journal} {\bibinfo  {journal} {Nature}\ }\textbf {\bibinfo
  {volume} {560}},\ \bibinfo {pages} {456} (\bibinfo {year}
  {2018})}\BibitemShut {NoStop}%
\bibitem [{\citenamefont {King}\ \emph {et~al.}(2023)\citenamefont {King},
  \citenamefont {Raymond}, \citenamefont {Lanting}, \citenamefont {Harris},
  \citenamefont {Zucca}, \citenamefont {Altomare}, \citenamefont {Berkley},
  \citenamefont {Boothby}, \citenamefont {Ejtemaee}, \citenamefont {Enderud}
  \emph {et~al.}}]{king2023quantum}%
  \BibitemOpen
  \bibfield  {author} {\bibinfo {author} {\bibfnamefont {A.~D.}\ \bibnamefont
  {King}}, \bibinfo {author} {\bibfnamefont {J.}~\bibnamefont {Raymond}},
  \bibinfo {author} {\bibfnamefont {T.}~\bibnamefont {Lanting}}, \bibinfo
  {author} {\bibfnamefont {R.}~\bibnamefont {Harris}}, \bibinfo {author}
  {\bibfnamefont {A.}~\bibnamefont {Zucca}}, \bibinfo {author} {\bibfnamefont
  {F.}~\bibnamefont {Altomare}}, \bibinfo {author} {\bibfnamefont {A.~J.}\
  \bibnamefont {Berkley}}, \bibinfo {author} {\bibfnamefont {K.}~\bibnamefont
  {Boothby}}, \bibinfo {author} {\bibfnamefont {S.}~\bibnamefont {Ejtemaee}},
  \bibinfo {author} {\bibfnamefont {C.}~\bibnamefont {Enderud}}, \emph
  {et~al.},\ }\bibfield  {title} {\bibinfo {title} {Quantum critical dynamics
  in a 5,000-qubit programmable spin glass},\ }\href@noop {} {\bibfield
  {journal} {\bibinfo  {journal} {Nature}\ ,\ \bibinfo {pages} {1}} (\bibinfo
  {year} {2023})}\BibitemShut {NoStop}%
\bibitem [{\citenamefont {Kato}(1950)}]{kato1950adiabatic}%
  \BibitemOpen
  \bibfield  {author} {\bibinfo {author} {\bibfnamefont {T.}~\bibnamefont
  {Kato}},\ }\bibfield  {title} {\bibinfo {title} {On the adiabatic theorem of
  quantum mechanics},\ }\href@noop {} {\bibfield  {journal} {\bibinfo
  {journal} {Journal of the Physical Society of Japan}\ }\textbf {\bibinfo
  {volume} {5}},\ \bibinfo {pages} {435} (\bibinfo {year} {1950})}\BibitemShut
  {NoStop}%
\bibitem [{\citenamefont {Messiah}(2014)}]{messiah2014quantum}%
  \BibitemOpen
  \bibfield  {author} {\bibinfo {author} {\bibfnamefont {A.}~\bibnamefont
  {Messiah}},\ }\href@noop {} {\emph {\bibinfo {title} {Quantum mechanics}}}\
  (\bibinfo  {publisher} {Courier Corporation},\ \bibinfo {year}
  {2014})\BibitemShut {NoStop}%
\bibitem [{\citenamefont {Jansen}\ \emph {et~al.}(2007)\citenamefont {Jansen},
  \citenamefont {Ruskai},\ and\ \citenamefont {Seiler}}]{jansen2007bounds}%
  \BibitemOpen
  \bibfield  {author} {\bibinfo {author} {\bibfnamefont {S.}~\bibnamefont
  {Jansen}}, \bibinfo {author} {\bibfnamefont {M.-B.}\ \bibnamefont {Ruskai}},\
  and\ \bibinfo {author} {\bibfnamefont {R.}~\bibnamefont {Seiler}},\
  }\bibfield  {title} {\bibinfo {title} {Bounds for the adiabatic approximation
  with applications to quantum computation},\ }\href@noop {} {\bibfield
  {journal} {\bibinfo  {journal} {Journal of Mathematical Physics}\ }\textbf
  {\bibinfo {volume} {48}} (\bibinfo {year} {2007})}\BibitemShut {NoStop}%
\bibitem [{\citenamefont {Aharonov}\ \emph {et~al.}(2008)\citenamefont
  {Aharonov}, \citenamefont {Van~Dam}, \citenamefont {Kempe}, \citenamefont
  {Landau}, \citenamefont {Lloyd},\ and\ \citenamefont
  {Regev}}]{aharonov2008adiabatic}%
  \BibitemOpen
  \bibfield  {author} {\bibinfo {author} {\bibfnamefont {D.}~\bibnamefont
  {Aharonov}}, \bibinfo {author} {\bibfnamefont {W.}~\bibnamefont {Van~Dam}},
  \bibinfo {author} {\bibfnamefont {J.}~\bibnamefont {Kempe}}, \bibinfo
  {author} {\bibfnamefont {Z.}~\bibnamefont {Landau}}, \bibinfo {author}
  {\bibfnamefont {S.}~\bibnamefont {Lloyd}},\ and\ \bibinfo {author}
  {\bibfnamefont {O.}~\bibnamefont {Regev}},\ }\bibfield  {title} {\bibinfo
  {title} {Adiabatic quantum computation is equivalent to standard quantum
  computation},\ }\href@noop {} {\bibfield  {journal} {\bibinfo  {journal}
  {SIAM review}\ }\textbf {\bibinfo {volume} {50}},\ \bibinfo {pages} {755}
  (\bibinfo {year} {2008})}\BibitemShut {NoStop}%
\bibitem [{\citenamefont {Biamonte}\ and\ \citenamefont
  {Love}(2008)}]{biamonte2008realizable}%
  \BibitemOpen
  \bibfield  {author} {\bibinfo {author} {\bibfnamefont {J.~D.}\ \bibnamefont
  {Biamonte}}\ and\ \bibinfo {author} {\bibfnamefont {P.~J.}\ \bibnamefont
  {Love}},\ }\bibfield  {title} {\bibinfo {title} {Realizable hamiltonians for
  universal adiabatic quantum computers},\ }\href@noop {} {\bibfield  {journal}
  {\bibinfo  {journal} {Physical Review A}\ }\textbf {\bibinfo {volume} {78}},\
  \bibinfo {pages} {012352} (\bibinfo {year} {2008})}\BibitemShut {NoStop}%
\bibitem [{\citenamefont {Mooij}\ \emph {et~al.}(1999)\citenamefont {Mooij},
  \citenamefont {Orlando}, \citenamefont {Levitov}, \citenamefont {Tian},
  \citenamefont {Van~der Wal},\ and\ \citenamefont
  {Lloyd}}]{mooij1999josephson}%
  \BibitemOpen
  \bibfield  {author} {\bibinfo {author} {\bibfnamefont {J.}~\bibnamefont
  {Mooij}}, \bibinfo {author} {\bibfnamefont {T.}~\bibnamefont {Orlando}},
  \bibinfo {author} {\bibfnamefont {L.}~\bibnamefont {Levitov}}, \bibinfo
  {author} {\bibfnamefont {L.}~\bibnamefont {Tian}}, \bibinfo {author}
  {\bibfnamefont {C.~H.}\ \bibnamefont {Van~der Wal}},\ and\ \bibinfo {author}
  {\bibfnamefont {S.}~\bibnamefont {Lloyd}},\ }\bibfield  {title} {\bibinfo
  {title} {Josephson persistent-current qubit},\ }\href@noop {} {\bibfield
  {journal} {\bibinfo  {journal} {Science}\ }\textbf {\bibinfo {volume}
  {285}},\ \bibinfo {pages} {1036} (\bibinfo {year} {1999})}\BibitemShut
  {NoStop}%
\bibitem [{\citenamefont {Orlando}\ \emph {et~al.}(1999)\citenamefont
  {Orlando}, \citenamefont {Mooij}, \citenamefont {Tian}, \citenamefont {Van
  Der~Wal}, \citenamefont {Levitov}, \citenamefont {Lloyd},\ and\ \citenamefont
  {Mazo}}]{orlando1999superconducting}%
  \BibitemOpen
  \bibfield  {author} {\bibinfo {author} {\bibfnamefont {T.}~\bibnamefont
  {Orlando}}, \bibinfo {author} {\bibfnamefont {J.}~\bibnamefont {Mooij}},
  \bibinfo {author} {\bibfnamefont {L.}~\bibnamefont {Tian}}, \bibinfo {author}
  {\bibfnamefont {C.~H.}\ \bibnamefont {Van Der~Wal}}, \bibinfo {author}
  {\bibfnamefont {L.}~\bibnamefont {Levitov}}, \bibinfo {author} {\bibfnamefont
  {S.}~\bibnamefont {Lloyd}},\ and\ \bibinfo {author} {\bibfnamefont
  {J.}~\bibnamefont {Mazo}},\ }\bibfield  {title} {\bibinfo {title}
  {Superconducting persistent-current qubit},\ }\href@noop {} {\bibfield
  {journal} {\bibinfo  {journal} {Physical Review B}\ }\textbf {\bibinfo
  {volume} {60}},\ \bibinfo {pages} {15398} (\bibinfo {year}
  {1999})}\BibitemShut {NoStop}%
\bibitem [{\citenamefont {Clarke}\ and\ \citenamefont
  {Wilhelm}(2008)}]{clarke2008superconducting}%
  \BibitemOpen
  \bibfield  {author} {\bibinfo {author} {\bibfnamefont {J.}~\bibnamefont
  {Clarke}}\ and\ \bibinfo {author} {\bibfnamefont {F.~K.}\ \bibnamefont
  {Wilhelm}},\ }\bibfield  {title} {\bibinfo {title} {Superconducting quantum
  bits},\ }\href@noop {} {\bibfield  {journal} {\bibinfo  {journal} {Nature}\
  }\textbf {\bibinfo {volume} {453}},\ \bibinfo {pages} {1031} (\bibinfo {year}
  {2008})}\BibitemShut {NoStop}%
\bibitem [{\citenamefont {Makhlin}\ \emph {et~al.}(2001)\citenamefont
  {Makhlin}, \citenamefont {Sch{\"o}n},\ and\ \citenamefont
  {Shnirman}}]{makhlin2001quantum}%
  \BibitemOpen
  \bibfield  {author} {\bibinfo {author} {\bibfnamefont {Y.}~\bibnamefont
  {Makhlin}}, \bibinfo {author} {\bibfnamefont {G.}~\bibnamefont {Sch{\"o}n}},\
  and\ \bibinfo {author} {\bibfnamefont {A.}~\bibnamefont {Shnirman}},\
  }\bibfield  {title} {\bibinfo {title} {Quantum-state engineering with
  josephson-junction devices},\ }\href@noop {} {\bibfield  {journal} {\bibinfo
  {journal} {Reviews of modern physics}\ }\textbf {\bibinfo {volume} {73}},\
  \bibinfo {pages} {357} (\bibinfo {year} {2001})}\BibitemShut {NoStop}%
\bibitem [{\citenamefont {Levine}\ \emph {et~al.}(2019)\citenamefont {Levine},
  \citenamefont {Keesling}, \citenamefont {Semeghini}, \citenamefont {Omran},
  \citenamefont {Wang}, \citenamefont {Ebadi}, \citenamefont {Bernien},
  \citenamefont {Greiner}, \citenamefont {Vuleti{\'c}}, \citenamefont {Pichler}
  \emph {et~al.}}]{levine2019parallel}%
  \BibitemOpen
  \bibfield  {author} {\bibinfo {author} {\bibfnamefont {H.}~\bibnamefont
  {Levine}}, \bibinfo {author} {\bibfnamefont {A.}~\bibnamefont {Keesling}},
  \bibinfo {author} {\bibfnamefont {G.}~\bibnamefont {Semeghini}}, \bibinfo
  {author} {\bibfnamefont {A.}~\bibnamefont {Omran}}, \bibinfo {author}
  {\bibfnamefont {T.~T.}\ \bibnamefont {Wang}}, \bibinfo {author}
  {\bibfnamefont {S.}~\bibnamefont {Ebadi}}, \bibinfo {author} {\bibfnamefont
  {H.}~\bibnamefont {Bernien}}, \bibinfo {author} {\bibfnamefont
  {M.}~\bibnamefont {Greiner}}, \bibinfo {author} {\bibfnamefont
  {V.}~\bibnamefont {Vuleti{\'c}}}, \bibinfo {author} {\bibfnamefont
  {H.}~\bibnamefont {Pichler}}, \emph {et~al.},\ }\bibfield  {title} {\bibinfo
  {title} {Parallel implementation of high-fidelity multiqubit gates with
  neutral atoms},\ }\href@noop {} {\bibfield  {journal} {\bibinfo  {journal}
  {Physical review letters}\ }\textbf {\bibinfo {volume} {123}},\ \bibinfo
  {pages} {170503} (\bibinfo {year} {2019})}\BibitemShut {NoStop}%
\bibitem [{\citenamefont {Chew}\ \emph {et~al.}(2022)\citenamefont {Chew},
  \citenamefont {Tomita}, \citenamefont {Mahesh}, \citenamefont {Sugawa},
  \citenamefont {de~L{\'e}s{\'e}leuc},\ and\ \citenamefont
  {Ohmori}}]{chew2022ultrafast}%
  \BibitemOpen
  \bibfield  {author} {\bibinfo {author} {\bibfnamefont {Y.}~\bibnamefont
  {Chew}}, \bibinfo {author} {\bibfnamefont {T.}~\bibnamefont {Tomita}},
  \bibinfo {author} {\bibfnamefont {T.~P.}\ \bibnamefont {Mahesh}}, \bibinfo
  {author} {\bibfnamefont {S.}~\bibnamefont {Sugawa}}, \bibinfo {author}
  {\bibfnamefont {S.}~\bibnamefont {de~L{\'e}s{\'e}leuc}},\ and\ \bibinfo
  {author} {\bibfnamefont {K.}~\bibnamefont {Ohmori}},\ }\bibfield  {title}
  {\bibinfo {title} {Ultrafast energy exchange between two single rydberg atoms
  on a nanosecond timescale},\ }\href@noop {} {\bibfield  {journal} {\bibinfo
  {journal} {Nature Photonics}\ }\textbf {\bibinfo {volume} {16}},\ \bibinfo
  {pages} {724} (\bibinfo {year} {2022})}\BibitemShut {NoStop}%
\bibitem [{\citenamefont {Bluvstein}\ \emph {et~al.}(2023)\citenamefont
  {Bluvstein}, \citenamefont {Evered}, \citenamefont {Geim}, \citenamefont
  {Li}, \citenamefont {Zhou}, \citenamefont {Manovitz}, \citenamefont {Ebadi},
  \citenamefont {Cain}, \citenamefont {Kalinowski}, \citenamefont {Hangleiter}
  \emph {et~al.}}]{bluvstein2023logical}%
  \BibitemOpen
  \bibfield  {author} {\bibinfo {author} {\bibfnamefont {D.}~\bibnamefont
  {Bluvstein}}, \bibinfo {author} {\bibfnamefont {S.~J.}\ \bibnamefont
  {Evered}}, \bibinfo {author} {\bibfnamefont {A.~A.}\ \bibnamefont {Geim}},
  \bibinfo {author} {\bibfnamefont {S.~H.}\ \bibnamefont {Li}}, \bibinfo
  {author} {\bibfnamefont {H.}~\bibnamefont {Zhou}}, \bibinfo {author}
  {\bibfnamefont {T.}~\bibnamefont {Manovitz}}, \bibinfo {author}
  {\bibfnamefont {S.}~\bibnamefont {Ebadi}}, \bibinfo {author} {\bibfnamefont
  {M.}~\bibnamefont {Cain}}, \bibinfo {author} {\bibfnamefont {M.}~\bibnamefont
  {Kalinowski}}, \bibinfo {author} {\bibfnamefont {D.}~\bibnamefont
  {Hangleiter}}, \emph {et~al.},\ }\bibfield  {title} {\bibinfo {title}
  {Logical quantum processor based on reconfigurable atom arrays},\ }\href@noop
  {} {\bibfield  {journal} {\bibinfo  {journal} {Nature}\ ,\ \bibinfo {pages}
  {1}} (\bibinfo {year} {2023})}\BibitemShut {NoStop}%
\bibitem [{\citenamefont {Imoto}\ \emph
  {et~al.}(2022{\natexlab{a}})\citenamefont {Imoto}, \citenamefont {Seki},\
  and\ \citenamefont {Matsuzaki}}]{imoto2022obtaining}%
  \BibitemOpen
  \bibfield  {author} {\bibinfo {author} {\bibfnamefont {T.}~\bibnamefont
  {Imoto}}, \bibinfo {author} {\bibfnamefont {Y.}~\bibnamefont {Seki}},\ and\
  \bibinfo {author} {\bibfnamefont {Y.}~\bibnamefont {Matsuzaki}},\ }\bibfield
  {title} {\bibinfo {title} {Obtaining ground states of the xxz model using the
  quantum annealing with inductively coupled superconducting flux qubits},\
  }\href@noop {} {\bibfield  {journal} {\bibinfo  {journal} {Journal of the
  Physical Society of Japan}\ }\textbf {\bibinfo {volume} {91}},\ \bibinfo
  {pages} {064004} (\bibinfo {year} {2022}{\natexlab{a}})}\BibitemShut
  {NoStop}%
\bibitem [{\citenamefont {Imoto}\ \emph
  {et~al.}(2022{\natexlab{b}})\citenamefont {Imoto}, \citenamefont {Seki},\
  and\ \citenamefont {Matsuzaki}}]{imoto2022quantum}%
  \BibitemOpen
  \bibfield  {author} {\bibinfo {author} {\bibfnamefont {T.}~\bibnamefont
  {Imoto}}, \bibinfo {author} {\bibfnamefont {Y.}~\bibnamefont {Seki}},\ and\
  \bibinfo {author} {\bibfnamefont {Y.}~\bibnamefont {Matsuzaki}},\ }\bibfield
  {title} {\bibinfo {title} {Quantum annealing with symmetric subspaces},\
  }\href@noop {} {\bibfield  {journal} {\bibinfo  {journal} {arXiv preprint
  arXiv:2209.09575}\ } (\bibinfo {year} {2022}{\natexlab{b}})}\BibitemShut
  {NoStop}%
\bibitem [{\citenamefont {Hatomura}\ \emph {et~al.}(2022)\citenamefont
  {Hatomura}, \citenamefont {Yoshinaga}, \citenamefont {Matsuzaki},\ and\
  \citenamefont {Tatsuta}}]{hatomura2022quantum}%
  \BibitemOpen
  \bibfield  {author} {\bibinfo {author} {\bibfnamefont {T.}~\bibnamefont
  {Hatomura}}, \bibinfo {author} {\bibfnamefont {A.}~\bibnamefont {Yoshinaga}},
  \bibinfo {author} {\bibfnamefont {Y.}~\bibnamefont {Matsuzaki}},\ and\
  \bibinfo {author} {\bibfnamefont {M.}~\bibnamefont {Tatsuta}},\ }\bibfield
  {title} {\bibinfo {title} {Quantum metrology based on symmetry-protected
  adiabatic transformation: imperfection, finite time duration, and
  dephasing},\ }\href@noop {} {\bibfield  {journal} {\bibinfo  {journal} {New
  Journal of Physics}\ }\textbf {\bibinfo {volume} {24}},\ \bibinfo {pages}
  {033005} (\bibinfo {year} {2022})}\BibitemShut {NoStop}%
\bibitem [{\citenamefont {Matsuzaki}\ \emph {et~al.}(2022)\citenamefont
  {Matsuzaki}, \citenamefont {Imoto},\ and\ \citenamefont
  {Susa}}]{matsuzaki2022generation}%
  \BibitemOpen
  \bibfield  {author} {\bibinfo {author} {\bibfnamefont {Y.}~\bibnamefont
  {Matsuzaki}}, \bibinfo {author} {\bibfnamefont {T.}~\bibnamefont {Imoto}},\
  and\ \bibinfo {author} {\bibfnamefont {Y.}~\bibnamefont {Susa}},\ }\bibfield
  {title} {\bibinfo {title} {Generation of multipartite entanglement between
  spin-1 particles with bifurcation-based quantum annealing},\ }\href@noop {}
  {\bibfield  {journal} {\bibinfo  {journal} {Scientific Reports}\ }\textbf
  {\bibinfo {volume} {12}},\ \bibinfo {pages} {14964} (\bibinfo {year}
  {2022})}\BibitemShut {NoStop}%
\bibitem [{\citenamefont {Pudenz}\ \emph {et~al.}(2014)\citenamefont {Pudenz},
  \citenamefont {Albash},\ and\ \citenamefont {Lidar}}]{pudenz2014error}%
  \BibitemOpen
  \bibfield  {author} {\bibinfo {author} {\bibfnamefont {K.~L.}\ \bibnamefont
  {Pudenz}}, \bibinfo {author} {\bibfnamefont {T.}~\bibnamefont {Albash}},\
  and\ \bibinfo {author} {\bibfnamefont {D.~A.}\ \bibnamefont {Lidar}},\
  }\bibfield  {title} {\bibinfo {title} {Error-corrected quantum annealing with
  hundreds of qubits},\ }\href@noop {} {\bibfield  {journal} {\bibinfo
  {journal} {Nature communications}\ }\textbf {\bibinfo {volume} {5}},\
  \bibinfo {pages} {3243} (\bibinfo {year} {2014})}\BibitemShut {NoStop}%
\bibitem [{\citenamefont {Lidar}\ \emph {et~al.}(1998)\citenamefont {Lidar},
  \citenamefont {Chuang},\ and\ \citenamefont {Whaley}}]{lidar1998decoherence}%
  \BibitemOpen
  \bibfield  {author} {\bibinfo {author} {\bibfnamefont {D.~A.}\ \bibnamefont
  {Lidar}}, \bibinfo {author} {\bibfnamefont {I.~L.}\ \bibnamefont {Chuang}},\
  and\ \bibinfo {author} {\bibfnamefont {K.~B.}\ \bibnamefont {Whaley}},\
  }\bibfield  {title} {\bibinfo {title} {Decoherence-free subspaces for quantum
  computation},\ }\href@noop {} {\bibfield  {journal} {\bibinfo  {journal}
  {Physical Review Letters}\ }\textbf {\bibinfo {volume} {81}},\ \bibinfo
  {pages} {2594} (\bibinfo {year} {1998})}\BibitemShut {NoStop}%
\bibitem [{\citenamefont {Preskill}(2018)}]{preskill2018quantum}%
  \BibitemOpen
  \bibfield  {author} {\bibinfo {author} {\bibfnamefont {J.}~\bibnamefont
  {Preskill}},\ }\bibfield  {title} {\bibinfo {title} {Quantum computing in the
  nisq era and beyond},\ }\href@noop {} {\bibfield  {journal} {\bibinfo
  {journal} {Quantum}\ }\textbf {\bibinfo {volume} {2}},\ \bibinfo {pages} {79}
  (\bibinfo {year} {2018})}\BibitemShut {NoStop}%
\bibitem [{\citenamefont {Peruzzo}\ \emph {et~al.}(2014)\citenamefont
  {Peruzzo}, \citenamefont {McClean}, \citenamefont {Shadbolt}, \citenamefont
  {Yung}, \citenamefont {Zhou}, \citenamefont {Love}, \citenamefont
  {Aspuru-Guzik},\ and\ \citenamefont {O’brien}}]{peruzzo2014variational}%
  \BibitemOpen
  \bibfield  {author} {\bibinfo {author} {\bibfnamefont {A.}~\bibnamefont
  {Peruzzo}}, \bibinfo {author} {\bibfnamefont {J.}~\bibnamefont {McClean}},
  \bibinfo {author} {\bibfnamefont {P.}~\bibnamefont {Shadbolt}}, \bibinfo
  {author} {\bibfnamefont {M.-H.}\ \bibnamefont {Yung}}, \bibinfo {author}
  {\bibfnamefont {X.-Q.}\ \bibnamefont {Zhou}}, \bibinfo {author}
  {\bibfnamefont {P.~J.}\ \bibnamefont {Love}}, \bibinfo {author}
  {\bibfnamefont {A.}~\bibnamefont {Aspuru-Guzik}},\ and\ \bibinfo {author}
  {\bibfnamefont {J.~L.}\ \bibnamefont {O’brien}},\ }\bibfield  {title}
  {\bibinfo {title} {A variational eigenvalue solver on a photonic quantum
  processor},\ }\href@noop {} {\bibfield  {journal} {\bibinfo  {journal}
  {Nature communications}\ }\textbf {\bibinfo {volume} {5}},\ \bibinfo {pages}
  {4213} (\bibinfo {year} {2014})}\BibitemShut {NoStop}%
\bibitem [{\citenamefont {Seki}\ and\ \citenamefont
  {Nishimori}(2012)}]{seki2012quantum}%
  \BibitemOpen
  \bibfield  {author} {\bibinfo {author} {\bibfnamefont {Y.}~\bibnamefont
  {Seki}}\ and\ \bibinfo {author} {\bibfnamefont {H.}~\bibnamefont
  {Nishimori}},\ }\bibfield  {title} {\bibinfo {title} {Quantum annealing with
  antiferromagnetic fluctuations},\ }\href@noop {} {\bibfield  {journal}
  {\bibinfo  {journal} {Physical Review E}\ }\textbf {\bibinfo {volume} {85}},\
  \bibinfo {pages} {051112} (\bibinfo {year} {2012})}\BibitemShut {NoStop}%
\bibitem [{\citenamefont {Seki}\ and\ \citenamefont
  {Nishimori}(2015)}]{seki2015quantum}%
  \BibitemOpen
  \bibfield  {author} {\bibinfo {author} {\bibfnamefont {Y.}~\bibnamefont
  {Seki}}\ and\ \bibinfo {author} {\bibfnamefont {H.}~\bibnamefont
  {Nishimori}},\ }\bibfield  {title} {\bibinfo {title} {Quantum annealing with
  antiferromagnetic transverse interactions for the hopfield model},\
  }\href@noop {} {\bibfield  {journal} {\bibinfo  {journal} {Journal of Physics
  A: Mathematical and Theoretical}\ }\textbf {\bibinfo {volume} {48}},\
  \bibinfo {pages} {335301} (\bibinfo {year} {2015})}\BibitemShut {NoStop}%
\bibitem [{\citenamefont {Takada}\ \emph {et~al.}(2020)\citenamefont {Takada},
  \citenamefont {Yamashiro},\ and\ \citenamefont {Nishimori}}]{takada2020mean}%
  \BibitemOpen
  \bibfield  {author} {\bibinfo {author} {\bibfnamefont {K.}~\bibnamefont
  {Takada}}, \bibinfo {author} {\bibfnamefont {Y.}~\bibnamefont {Yamashiro}},\
  and\ \bibinfo {author} {\bibfnamefont {H.}~\bibnamefont {Nishimori}},\
  }\bibfield  {title} {\bibinfo {title} {Mean-field solution of the weak-strong
  cluster problem for quantum annealing with stoquastic and non-stoquastic
  catalysts},\ }\href@noop {} {\bibfield  {journal} {\bibinfo  {journal}
  {Journal of the Physical Society of Japan}\ }\textbf {\bibinfo {volume}
  {89}},\ \bibinfo {pages} {044001} (\bibinfo {year} {2020})}\BibitemShut
  {NoStop}%
\bibitem [{\citenamefont {Susa}\ \emph {et~al.}(2023)\citenamefont {Susa},
  \citenamefont {Imoto},\ and\ \citenamefont
  {Matsuzaki}}]{susa2023nonstoquastic}%
  \BibitemOpen
  \bibfield  {author} {\bibinfo {author} {\bibfnamefont {Y.}~\bibnamefont
  {Susa}}, \bibinfo {author} {\bibfnamefont {T.}~\bibnamefont {Imoto}},\ and\
  \bibinfo {author} {\bibfnamefont {Y.}~\bibnamefont {Matsuzaki}},\ }\bibfield
  {title} {\bibinfo {title} {Nonstoquastic catalyst for bifurcation-based
  quantum annealing of the ferromagnetic p-spin model},\ }\href@noop {}
  {\bibfield  {journal} {\bibinfo  {journal} {Physical Review A}\ }\textbf
  {\bibinfo {volume} {107}},\ \bibinfo {pages} {052401} (\bibinfo {year}
  {2023})}\BibitemShut {NoStop}%
\bibitem [{\citenamefont {Berry}(2009)}]{berry2009transitionless}%
  \BibitemOpen
  \bibfield  {author} {\bibinfo {author} {\bibfnamefont {M.~V.}\ \bibnamefont
  {Berry}},\ }\bibfield  {title} {\bibinfo {title} {Transitionless quantum
  driving},\ }\href@noop {} {\bibfield  {journal} {\bibinfo  {journal} {Journal
  of Physics A: Mathematical and Theoretical}\ }\textbf {\bibinfo {volume}
  {42}},\ \bibinfo {pages} {365303} (\bibinfo {year} {2009})}\BibitemShut
  {NoStop}%
\bibitem [{\citenamefont {Torrontegui}\ \emph {et~al.}(2013)\citenamefont
  {Torrontegui}, \citenamefont {Ib{\'a}{\~n}ez}, \citenamefont
  {Mart{\'\i}nez-Garaot}, \citenamefont {Modugno}, \citenamefont {del Campo},
  \citenamefont {Gu{\'e}ry-Odelin}, \citenamefont {Ruschhaupt}, \citenamefont
  {Chen},\ and\ \citenamefont {Muga}}]{torrontegui2013shortcuts}%
  \BibitemOpen
  \bibfield  {author} {\bibinfo {author} {\bibfnamefont {E.}~\bibnamefont
  {Torrontegui}}, \bibinfo {author} {\bibfnamefont {S.}~\bibnamefont
  {Ib{\'a}{\~n}ez}}, \bibinfo {author} {\bibfnamefont {S.}~\bibnamefont
  {Mart{\'\i}nez-Garaot}}, \bibinfo {author} {\bibfnamefont {M.}~\bibnamefont
  {Modugno}}, \bibinfo {author} {\bibfnamefont {A.}~\bibnamefont {del Campo}},
  \bibinfo {author} {\bibfnamefont {D.}~\bibnamefont {Gu{\'e}ry-Odelin}},
  \bibinfo {author} {\bibfnamefont {A.}~\bibnamefont {Ruschhaupt}}, \bibinfo
  {author} {\bibfnamefont {X.}~\bibnamefont {Chen}},\ and\ \bibinfo {author}
  {\bibfnamefont {J.~G.}\ \bibnamefont {Muga}},\ }\bibfield  {title} {\bibinfo
  {title} {Shortcuts to adiabaticity},\ }in\ \href@noop {} {\emph {\bibinfo
  {booktitle} {Advances in atomic, molecular, and optical physics}}},\
  Vol.~\bibinfo {volume} {62}\ (\bibinfo  {publisher} {Elsevier},\ \bibinfo
  {year} {2013})\ pp.\ \bibinfo {pages} {117--169}\BibitemShut {NoStop}%
\bibitem [{\citenamefont {Gu{\'e}ry-Odelin}\ \emph {et~al.}(2019)\citenamefont
  {Gu{\'e}ry-Odelin}, \citenamefont {Ruschhaupt}, \citenamefont {Kiely},
  \citenamefont {Torrontegui}, \citenamefont {Mart{\'\i}nez-Garaot},\ and\
  \citenamefont {Muga}}]{guery2019shortcuts}%
  \BibitemOpen
  \bibfield  {author} {\bibinfo {author} {\bibfnamefont {D.}~\bibnamefont
  {Gu{\'e}ry-Odelin}}, \bibinfo {author} {\bibfnamefont {A.}~\bibnamefont
  {Ruschhaupt}}, \bibinfo {author} {\bibfnamefont {A.}~\bibnamefont {Kiely}},
  \bibinfo {author} {\bibfnamefont {E.}~\bibnamefont {Torrontegui}}, \bibinfo
  {author} {\bibfnamefont {S.}~\bibnamefont {Mart{\'\i}nez-Garaot}},\ and\
  \bibinfo {author} {\bibfnamefont {J.~G.}\ \bibnamefont {Muga}},\ }\bibfield
  {title} {\bibinfo {title} {Shortcuts to adiabaticity: Concepts, methods, and
  applications},\ }\href@noop {} {\bibfield  {journal} {\bibinfo  {journal}
  {Reviews of Modern Physics}\ }\textbf {\bibinfo {volume} {91}},\ \bibinfo
  {pages} {045001} (\bibinfo {year} {2019})}\BibitemShut {NoStop}%
\bibitem [{\citenamefont {Lovett}\ \emph {et~al.}(2010)\citenamefont {Lovett},
  \citenamefont {Cooper}, \citenamefont {Everitt}, \citenamefont {Trevers},\
  and\ \citenamefont {Kendon}}]{lovett2010universal}%
  \BibitemOpen
  \bibfield  {author} {\bibinfo {author} {\bibfnamefont {N.~B.}\ \bibnamefont
  {Lovett}}, \bibinfo {author} {\bibfnamefont {S.}~\bibnamefont {Cooper}},
  \bibinfo {author} {\bibfnamefont {M.}~\bibnamefont {Everitt}}, \bibinfo
  {author} {\bibfnamefont {M.}~\bibnamefont {Trevers}},\ and\ \bibinfo {author}
  {\bibfnamefont {V.}~\bibnamefont {Kendon}},\ }\bibfield  {title} {\bibinfo
  {title} {Universal quantum computation using the discrete-time quantum
  walk},\ }\href@noop {} {\bibfield  {journal} {\bibinfo  {journal} {Physical
  Review A}\ }\textbf {\bibinfo {volume} {81}},\ \bibinfo {pages} {042330}
  (\bibinfo {year} {2010})}\BibitemShut {NoStop}%
\bibitem [{\citenamefont {Childs}(2009)}]{childs2009universal}%
  \BibitemOpen
  \bibfield  {author} {\bibinfo {author} {\bibfnamefont {A.~M.}\ \bibnamefont
  {Childs}},\ }\bibfield  {title} {\bibinfo {title} {Universal computation by
  quantum walk},\ }\href@noop {} {\bibfield  {journal} {\bibinfo  {journal}
  {Physical review letters}\ }\textbf {\bibinfo {volume} {102}},\ \bibinfo
  {pages} {180501} (\bibinfo {year} {2009})}\BibitemShut {NoStop}%
\bibitem [{\citenamefont {Childs}\ \emph {et~al.}(2003)\citenamefont {Childs},
  \citenamefont {Cleve}, \citenamefont {Deotto}, \citenamefont {Farhi},
  \citenamefont {Gutmann},\ and\ \citenamefont
  {Spielman}}]{childs2003exponential}%
  \BibitemOpen
  \bibfield  {author} {\bibinfo {author} {\bibfnamefont {A.~M.}\ \bibnamefont
  {Childs}}, \bibinfo {author} {\bibfnamefont {R.}~\bibnamefont {Cleve}},
  \bibinfo {author} {\bibfnamefont {E.}~\bibnamefont {Deotto}}, \bibinfo
  {author} {\bibfnamefont {E.}~\bibnamefont {Farhi}}, \bibinfo {author}
  {\bibfnamefont {S.}~\bibnamefont {Gutmann}},\ and\ \bibinfo {author}
  {\bibfnamefont {D.~A.}\ \bibnamefont {Spielman}},\ }\bibfield  {title}
  {\bibinfo {title} {Exponential algorithmic speedup by a quantum walk},\ }in\
  \href@noop {} {\emph {\bibinfo {booktitle} {Proceedings of the thirty-fifth
  annual ACM symposium on Theory of computing}}}\ (\bibinfo {year} {2003})\
  pp.\ \bibinfo {pages} {59--68}\BibitemShut {NoStop}%
\bibitem [{\citenamefont {Childs}\ \emph {et~al.}(2013)\citenamefont {Childs},
  \citenamefont {Gosset},\ and\ \citenamefont {Webb}}]{childs2013universal}%
  \BibitemOpen
  \bibfield  {author} {\bibinfo {author} {\bibfnamefont {A.~M.}\ \bibnamefont
  {Childs}}, \bibinfo {author} {\bibfnamefont {D.}~\bibnamefont {Gosset}},\
  and\ \bibinfo {author} {\bibfnamefont {Z.}~\bibnamefont {Webb}},\ }\bibfield
  {title} {\bibinfo {title} {Universal computation by multiparticle quantum
  walk},\ }\href@noop {} {\bibfield  {journal} {\bibinfo  {journal} {Science}\
  }\textbf {\bibinfo {volume} {339}},\ \bibinfo {pages} {791} (\bibinfo {year}
  {2013})}\BibitemShut {NoStop}%
\bibitem [{\citenamefont {Raussendorf}\ and\ \citenamefont
  {Briegel}(2001)}]{raussendorf2001one}%
  \BibitemOpen
  \bibfield  {author} {\bibinfo {author} {\bibfnamefont {R.}~\bibnamefont
  {Raussendorf}}\ and\ \bibinfo {author} {\bibfnamefont {H.~J.}\ \bibnamefont
  {Briegel}},\ }\bibfield  {title} {\bibinfo {title} {A one-way quantum
  computer},\ }\href@noop {} {\bibfield  {journal} {\bibinfo  {journal}
  {Physical review letters}\ }\textbf {\bibinfo {volume} {86}},\ \bibinfo
  {pages} {5188} (\bibinfo {year} {2001})}\BibitemShut {NoStop}%
\bibitem [{\citenamefont {Hen}(2015)}]{hen2015quantum}%
  \BibitemOpen
  \bibfield  {author} {\bibinfo {author} {\bibfnamefont {I.}~\bibnamefont
  {Hen}},\ }\bibfield  {title} {\bibinfo {title} {Quantum gates with controlled
  adiabatic evolutions},\ }\href@noop {} {\bibfield  {journal} {\bibinfo
  {journal} {Physical Review A}\ }\textbf {\bibinfo {volume} {91}},\ \bibinfo
  {pages} {022309} (\bibinfo {year} {2015})}\BibitemShut {NoStop}%
\bibitem [{\citenamefont {Bacon}\ and\ \citenamefont
  {Flammia}(2010)}]{bacon2010adiabatic}%
  \BibitemOpen
  \bibfield  {author} {\bibinfo {author} {\bibfnamefont {D.}~\bibnamefont
  {Bacon}}\ and\ \bibinfo {author} {\bibfnamefont {S.~T.}\ \bibnamefont
  {Flammia}},\ }\bibfield  {title} {\bibinfo {title} {Adiabatic cluster-state
  quantum computing},\ }\href@noop {} {\bibfield  {journal} {\bibinfo
  {journal} {Physical Review A}\ }\textbf {\bibinfo {volume} {82}},\ \bibinfo
  {pages} {030303} (\bibinfo {year} {2010})}\BibitemShut {NoStop}%
\bibitem [{\citenamefont {Ribeiro}\ and\ \citenamefont
  {Clerk}(2019)}]{ribeiro2019accelerated}%
  \BibitemOpen
  \bibfield  {author} {\bibinfo {author} {\bibfnamefont {H.}~\bibnamefont
  {Ribeiro}}\ and\ \bibinfo {author} {\bibfnamefont {A.~A.}\ \bibnamefont
  {Clerk}},\ }\bibfield  {title} {\bibinfo {title} {Accelerated adiabatic
  quantum gates: optimizing speed versus robustness},\ }\href@noop {}
  {\bibfield  {journal} {\bibinfo  {journal} {Physical Review A}\ }\textbf
  {\bibinfo {volume} {100}},\ \bibinfo {pages} {032323} (\bibinfo {year}
  {2019})}\BibitemShut {NoStop}%
\bibitem [{\citenamefont {Masuda}\ \emph {et~al.}(2022)\citenamefont {Masuda},
  \citenamefont {Kanao}, \citenamefont {Goto}, \citenamefont {Matsuzaki},
  \citenamefont {Ishikawa},\ and\ \citenamefont {Kawabata}}]{masuda2022fast}%
  \BibitemOpen
  \bibfield  {author} {\bibinfo {author} {\bibfnamefont {S.}~\bibnamefont
  {Masuda}}, \bibinfo {author} {\bibfnamefont {T.}~\bibnamefont {Kanao}},
  \bibinfo {author} {\bibfnamefont {H.}~\bibnamefont {Goto}}, \bibinfo {author}
  {\bibfnamefont {Y.}~\bibnamefont {Matsuzaki}}, \bibinfo {author}
  {\bibfnamefont {T.}~\bibnamefont {Ishikawa}},\ and\ \bibinfo {author}
  {\bibfnamefont {S.}~\bibnamefont {Kawabata}},\ }\bibfield  {title} {\bibinfo
  {title} {Fast tunable coupling scheme of kerr parametric oscillators based on
  shortcuts to adiabaticity},\ }\href@noop {} {\bibfield  {journal} {\bibinfo
  {journal} {Physical Review Applied}\ }\textbf {\bibinfo {volume} {18}},\
  \bibinfo {pages} {034076} (\bibinfo {year} {2022})}\BibitemShut {NoStop}%
\bibitem [{\citenamefont {Wang}\ \emph {et~al.}(2018)\citenamefont {Wang},
  \citenamefont {Zhang}, \citenamefont {Xiang}, \citenamefont {Jia},
  \citenamefont {Duan}, \citenamefont {Cai}, \citenamefont {Gong},
  \citenamefont {Zong}, \citenamefont {Wu}, \citenamefont {Wu} \emph
  {et~al.}}]{wang2018experimental}%
  \BibitemOpen
  \bibfield  {author} {\bibinfo {author} {\bibfnamefont {T.}~\bibnamefont
  {Wang}}, \bibinfo {author} {\bibfnamefont {Z.}~\bibnamefont {Zhang}},
  \bibinfo {author} {\bibfnamefont {L.}~\bibnamefont {Xiang}}, \bibinfo
  {author} {\bibfnamefont {Z.}~\bibnamefont {Jia}}, \bibinfo {author}
  {\bibfnamefont {P.}~\bibnamefont {Duan}}, \bibinfo {author} {\bibfnamefont
  {W.}~\bibnamefont {Cai}}, \bibinfo {author} {\bibfnamefont {Z.}~\bibnamefont
  {Gong}}, \bibinfo {author} {\bibfnamefont {Z.}~\bibnamefont {Zong}}, \bibinfo
  {author} {\bibfnamefont {M.}~\bibnamefont {Wu}}, \bibinfo {author}
  {\bibfnamefont {J.}~\bibnamefont {Wu}}, \emph {et~al.},\ }\bibfield  {title}
  {\bibinfo {title} {The experimental realization of high-fidelity
  ‘shortcut-to-adiabaticity’quantum gates in a superconducting xmon
  qubit},\ }\href@noop {} {\bibfield  {journal} {\bibinfo  {journal} {New
  Journal of Physics}\ }\textbf {\bibinfo {volume} {20}},\ \bibinfo {pages}
  {065003} (\bibinfo {year} {2018})}\BibitemShut {NoStop}%
\end{thebibliography}%

\appendix

\section{Creation of an arbitrary superposition adiabatically using degeneracy}\label{sec:superposition}

Here, we explain how to create an arbitrary superposition between degenerated ground states of the problem Hamiltonian. Although we explained a special case that the problem Hamiltonian is the identity in the main text, we will show that the concept introduced in the main text can be generalized to a broader class. Especially, we consider a case where the problem Hamiltonian is the Ising model. The prescription is as follows.
First, we adopt the Ising model whose ground states are degenerate and choose this as the problem Hamiltonian. Second, we perform QA where we add a catalytic term of the longitudinal field in the middle of the QA.
Finally, we obtain a superposition of the degenerate ground states after QA as long as the adiabatic condition is satisfied.
Importantly, if the longitudinal magnetic field is turned on and off during this annealing operation, the amplitude of the superposition between the ground states is controlled.
\begin{figure}[h!]
    \centering
    \includegraphics[width=90mm]{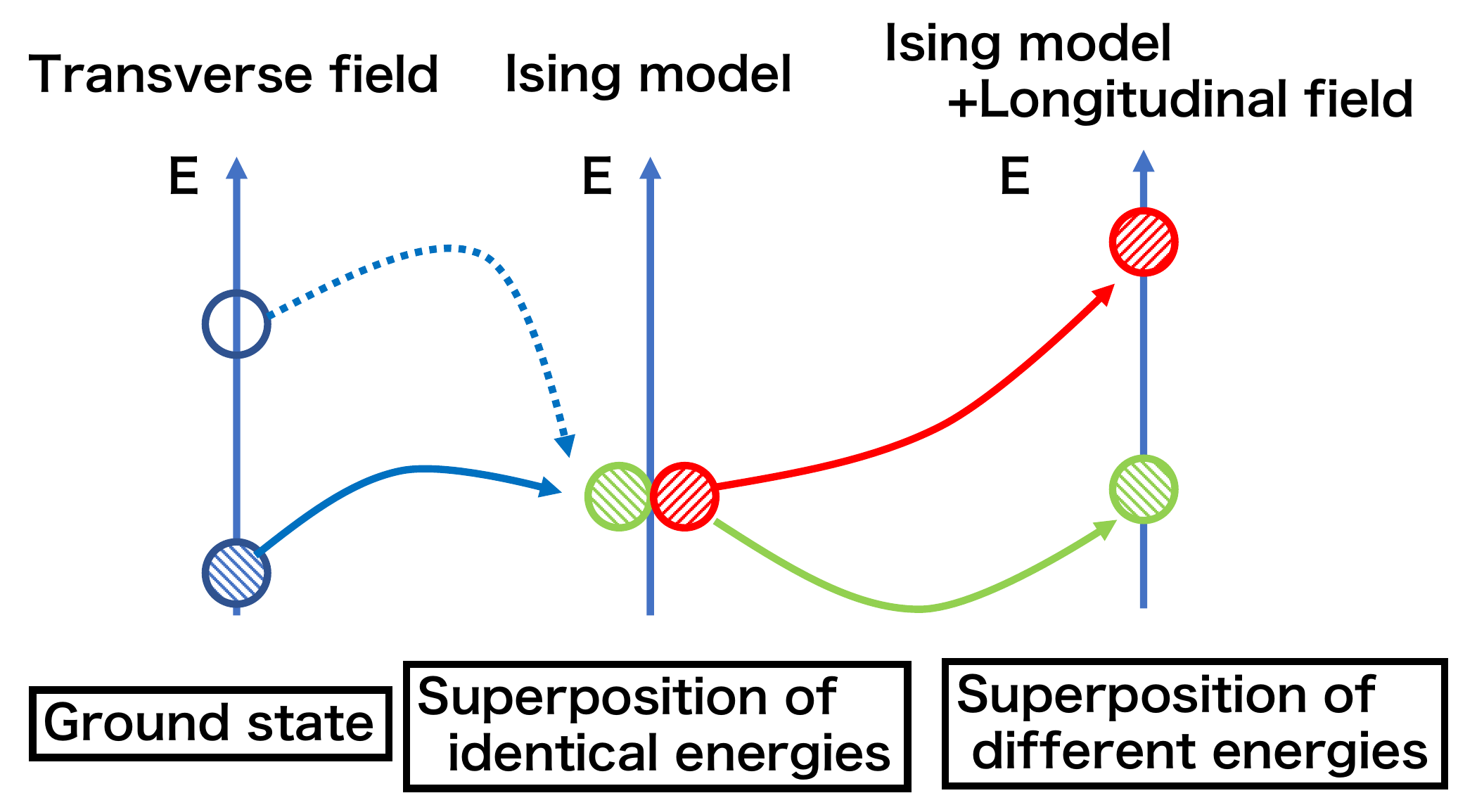}
    \caption{The concept of creation of the superpositioned state.
    We consider the transverse field as the drive Hamiltonian and, the degenerate Ising model as the problem Hamiltonian.
    Thus, we obtain the superposition of identical energies when the Hamiltonian is the Ising model.
    In addition, by adding the longitudinal magnetic field to this Hamiltonian, we obtain the superposition of different energies.
    }
    \label{fig:create_superposition}
\end{figure}
In addition, we introduce several Hamiltonian to have degenerate ground states in the following subsections. Our method can be implementable with the D-wave device.

\subsection{Two-state entangled states}\label{sec:two_state_entangled}
We consider the following Ising model as the problem Hamiltonian.
\begin{align}
    H=-\biggl(\sum_{i=1}^{N}\alpha_{i}\hat{\sigma}_{i}^{(z)}\biggr)^{2},\ \ \alpha_{i}\in\{+1, -1\}.
\end{align}
This Hamiltonian has the following two degenerated ground states.
\begin{align}
    \ket{\Phi}=J_{+}^{(x)}\ket{\downarrow \downarrow\cdots\downarrow},\ J_{-}^{(x)}\ket{\downarrow\downarrow, \cdots\downarrow} 
\end{align}
where $J_{+}^{(x)}=\prod_{\alpha_{j}=1}\hat{\sigma}_{j}^{(x)}$,\ $J_{-}^{(x)}=\prod_{\alpha_{j}=-1}\hat{\sigma}_{j}^{(x)}$.
Thus, we can obtain the arbitrary superposition between these two states, including entangled states, via QA as long as adiabatically.
For example, we consider the ferromagnetic all-to-all interaction as follows.
\begin{align}
    H=-\biggl(\sum_{i=1}^{N}\hat{\sigma}_{i}^{(z)}\biggr)^{2}.
\end{align}
where $\alpha _i=1$ for all $i$.
The ground states of this Hamiltonian are the following.
\begin{align}
    \ket{\uparrow\uparrow\cdots\uparrow}, \ket{\downarrow\downarrow\cdots\downarrow}.
\end{align}
Using our proposed method, we can create an arbitrary superposition between these two states.

\subsection{Two-state product states}\label{sec:two_state_product}

Also, our method is applicable to arbitrary two-state product states.
We set the following term as the problem Hamiltonian.

\begin{align}
    H=-\biggl(\sum_{i\in A}\alpha_{i}\hat{\sigma}_{i}^{(z)}\biggr)^{2}+\biggl(\sum_{i\in B}\alpha_{i}\hat{\sigma}_{i}^{(z)}\biggr),\ \ \alpha_{i}\in\{+1, -1\}
\end{align}
where $A$, $B$ are denoted by partition of the set $[1,2,\cdots N]$ i.e.
\begin{align}
    A\cap B=\emptyset, A\cup B=[1,2,\cdots,N].
\end{align}
This Hamiltonian has the following two degenerated ground states.
\begin{align}
    \ket{\Phi}=J_{A+}^{(x)}J_{B+}^{(x)}\ket{\downarrow\downarrow\cdots\downarrow},\ J_{A+}^{(x)}J_{B+}^{(x)}J_{A}^{(x)}\ket{\downarrow\downarrow\cdots\downarrow}
\end{align}
where $J_{A}^{(x)}=\prod_{j\in A}\hat{\sigma}_{j}^{(x)}$, 
$J_{A+}^{(x)}=\prod_{j\in A\land\alpha_{j}=-1}\hat{\sigma}_{j}^{(x)}$, and
$J_{B+}^{(x)}=\prod_{j\in B\land\alpha_{j}=-1}\hat{\sigma}_{j}^{(x)}$.
Thus, the qubits including the part of $B$ are the same spin states, and, those that include the part of $A$ are different spin states.

For example, we consider the following Hamiltonian as a problem Hamiltonian.
\begin{align}
    H=-\biggl(\sum_{i=1}^{N/3}\alpha_{i}\hat{\sigma}_{i}^{(z)}\biggr)^{2}+\biggl(\sum_{i=N/3+1}^{2N/3}\hat{\sigma}_{i}^{(z)}\biggr)+\biggl(-\sum_{i=2N/3+1}^{N}\hat{\sigma}_{i}^{(z)}\biggr).
\end{align}

The ground states of this Hamiltonian consist of the superposition of the following two states.

\begin{align}
   \ket{\uparrow\uparrow\cdots\uparrow}_{1: N/3}\ket{\downarrow\downarrow\cdots\downarrow}_{N/3+1:2N/3}\ket{\uparrow\uparrow\cdots\uparrow}_{2N/3+1:1}\\
   \ket{\downarrow\downarrow\cdots\downarrow}_{1: N/3}\ket{\uparrow\uparrow\cdots\uparrow}_{N/3+1:2N/3}\ket{\downarrow\downarrow\cdots\downarrow}_{2N/3+1:1}
\end{align}
where $\ket{\uparrow\uparrow\cdots\uparrow}_{a: b}$ is denoted by the state on the $[a, b]$.

Thus, we see that $1\sim N/3$ spins are different and $N/3+1\sim N$ spins are the same, and the ground state is two states that are degenerate; if in superposition, it is a product
state.

\subsection{The combination of two-state entangled states and product states}

We consider combining section \ref{sec:two_state_entangled} and section \ref{sec:two_state_product} to realize more than three superposition states.
We remark that the superposition state obtained by this subsection is a product state.
The Hamiltonian which we deal with in this subsection is
\begin{align}
    H=-\sum_{j=1}^{K}\Bigl(\sum_{i\in A_{j}}\alpha_{i}\hat{\sigma}_{i}^{(z)}\Bigr)^{2}-\biggl(\sum_{i\in A_{K+1}}\alpha_{i}\hat{\sigma}_{i}^{(z)}\biggr)
\end{align}
where $A_{i}\cap A_{j}=\emptyset$ for $i\neq j$, and $\cup_{j=1}^{K+1}A_{j}=[1,2,\cdots,K+1]$.
Here, $\alpha_{i}\in\{+1, -1\}$ is the real coefficient.
The degenerate ground states of this Hamiltonian are

\begin{align}    
J_{A_{K+1}+}^{(x)}\prod_{i=1}^{K}\biggl(a_{i}+\sqrt{1-a_{i}^{2}}J_{A_{i}}^{(x)}\biggr)J_{A_{i}+}^{(x)}\ket{\downarrow\downarrow\downarrow\cdots\downarrow}
\end{align}
where $J_{A_{i}}^{(x)}=\prod_{j\in A_{i}}\hat{\sigma}_{j}^{(x)}$, $J_{A_{i}+}^{(x)}=\prod_{j\in A_{i}\land\alpha_{j}=-1}\hat{\sigma}_{j}^{(x)}$, $J_{A_{K+1}+}^{(x)}=\prod_{j\in A_{K+1}\land\alpha_{j}=-1}\hat{\sigma}_{j}^{(x)}$ and, $a_i$ is amplitude of the all down state on the domain $A_{j}$.
We can rewrite this state as follows

\begin{align}
    \bigotimes_{j=1}^{K}J_{A_{j}+}^{(x)}\biggl(&a_{i}\ket{\downarrow\downarrow\cdots\downarrow}_{A_{j}}\notag\\
    &+\sqrt{1-a_{i}^{2}}\ket{\uparrow\uparrow\cdots\uparrow}_{A_{j}}\biggr)\otimes J_{A_{K+1}+}^{(x)}\ket{\downarrow\downarrow\cdots\downarrow}_{K+1}
\end{align}

There are $2^K$ fold degeneracies for these ground states. Actually, if we take $a_i=\pm 1$ for $1,2,\cdots,K$, these states are orthogonal to each other.
In this case, we choose $H_C$ to apply magnetic fields where the same magnetic fields are applied in the same domain, but different magnetic fields are applied in a different domain.

Although we discussed the case of the Ising model as the problem Hamiltonian, we can generalize this idea to the other model as well. As long as the problem Hamiltonian has the degenerate ground states, a similar discussion can be applied to the other model.
The details
are left to future work

\section{Perturbative analysis of our system}\label{sec:perturbative_theory}

In this appendix, we discuss how the degeneracy of our problem Hamiltonian is resolved by using the perturbative analysis. This elucidates the role of the longitudinal magnetic fields when we implement our method for the gate operations.


\subsection{First order perturbation theory formulation for degenerate Hamiltonian}

We consider the perturbed Hamiltonian as 
\begin{align}
    H=H_{0}+\lambda V
\end{align}
where $H_{0}$ is the unperturbed Hamiltonian, $V$ is the perturbation Hamiltonian, and $\lambda$ is the small perturbation parameter.
The perturbative expansion of the eigenenergy and eigenstate is defined by 
\begin{align}
    E_{n}&=E_{n}^{(0)}+\lambda E_{n}^{(1)}+\lambda^{2} E_{n}^{(2)}\cdots\label{eq:perturbative_energy}\\
    \ket{\phi_{n}}&=\ket{\phi_{n}^{(0)}}+\lambda\ket{\phi_{n}^{(1)}}+\lambda^{2}\ket{\phi_{n}^{(2)}}\label{eq:perturbative_ground_state}\cdots
\end{align}
where $E_{n}^{(k)}$ is denoted by the $k$-th order perturbation of the $n$-th level eigenenergy and $\ket{\phi_{n}^{(k)}}$ is denoted by $k$-th order perturbation of the $n$-th level eigenstate.
We remark that $E_{n}^{(0)}$ ($\ket{\phi_{n}^{0}}$) is eigenenergy(eigenstate) of the perturbed Hamiltonian $H_{0}$.

When the unperturbed Hamiltonian $H_{0}$ has a two-fold degenerate ground state, the $0$-th order eigenstates are given by

\begin{align}
    \ket{\phi_{0}^{(0)}}_{\pm}=c_{\pm,1}\ket{1}+c_{\pm,2}\ket{2}\label{eq:degenerated_gs}
\end{align}
where $\ket{1}$ and $\ket{2}$ are the ground states of the unperturbed Hamiltonian $H_{0}$ and $c_{\pm,1}$($c_{\pm,2}$) is the amplitude of the $\ket{1}(\ket{0})$.
We remark that $c_{+,1}^{*}c_{-,1}+c_{+,2}^{*}c_{-,2}=0$ from orthogonality of the two ground states $\ket{\phi_{0}^{(0)}}_{\pm}$.
Also, to satisfy the normalization of the eigenstate, we obtain $|c_{+,1}|^{2}+|c_{+,2}|^{2}=c_{-,1}^{2}+c_{-,2}^{2}=1$.
In this case, the perturbation energy formula is given by
\begin{align}
    E_{n,\pm}^{1}=\frac{1}{2}\biggl\{(V_{11}+V_{22})\pm\sqrt{(V_{11}-V_{22})^{2}+4|V_{12}|^{2}}\biggr\}\label{eq:pm_1st_order_perturbative_energy}
\end{align}
and the ratio of superposition is given by
\begin{align}
    \frac{c_{\pm,1}}{c_{\pm,2}}=\frac{V_{12}}{E_{1\pm}-V_{11}}=\frac{E_{1\pm}-V_{22}}{V_{21}}\label{eq:ratio_c1c2}
\end{align}
where $V_{nm}=\bra{n}V\ket{m},\ n,m=1,2$.
We note that to obtain the population of each degenerated state, we need only the coefficient of $0$-th order perturbated states (\ref{eq:degenerated_gs}).

In this case, we regard $H_{P}$ as the unperturbed Hamiltonian $H_{0}$, and $H_{D}+kH_{C}$ as the perturbed Hamiltonian $V$.
Also, we consider the eigenstates $\ket{1}$ and $\ket{2}$ in (\ref{eq:degenerated_gs}) as the two degenerated groundstates written by the computational basis.

\subsection{Application of the perturbative analysis to our method}

Let us substitute $H_{0}=H_{P}$ and $V=H_{D}+h_{z}H_{C}$ where $H_{P}$ is the problem Hamiltonian, $H_{D}$ is the drive Hamiltonian, and $H_{C}$ is the catalytic Hamiltonian in our method.
Also, we consider the eigenstates $\ket{1}$ and $\ket{2}$ in (\ref{eq:degenerated_gs}) as the two degenerated groundstates written by the computational basis.

\subsubsection{Perturbative analysis of X-rotation gate}\label{xrotationappendix}

First, we consider our method described in the section \ref{sec:realize_x_rotation}.
The two degenerated ground states are $\ket{1}$ and $\ket{2}$ as follows.
\begin{align}
    \ket{1}&=\ket{\uparrow}\label{eq:degenerated_ge_11}\\
    \ket{2}&=\ket{\downarrow}\label{eq:degenerated_ge_12}
\end{align}
We obtain the $V_{nm},\ n,m=1,2$ using the (\ref{eq:degenerated_ge_11}), (\ref{eq:degenerated_ge_12}) as follows.
\begin{align}
    V_{11}&=h_{z}\\
    V_{22}&=-h_{z}\\
    V_{12}&=-1\\
    V_{21}&=-1
\end{align}

In addition, the first-order perturbation energy $E_{0,\pm}^{(1)}$ in Eq. (\ref{eq:pm_1st_order_perturbative_energy}) is given by
\begin{align}
    E_{0,\pm}^{(1)}=\pm\sqrt{1+h_{z}^{2}}
\end{align}
The amplitude of the two degenerated ground states is given by
\begin{align}
    \frac{c_{\pm,1}}{c_{\pm,2}}=\frac{1}{h_{z}\mp\sqrt{1+h_{z}^{2}}}.
\end{align}
Using the normalization condition, we obtain the amplitude of $\ket{\uparrow}$ and $\ket{\downarrow}$ of the $0$-th order perturbated ground states as follows.

\begin{align}
    c_{\pm,1}^{2}&=\frac{1}{1+(\pm\sqrt{1+h_{z}^{2}}-h_{z})^{2}}\label{eq:single_x_rotation_analytic_1}.\\
    c_{\pm,2}^{2}&=\frac{(\pm\sqrt{1+h_{z}^{2}}-h_{z})^{2}}{1+(\pm\sqrt{1+h_{z}^{2}}-h_{z})^{2}}\label{eq:single_x_rotation_analytic_2}.
\end{align}

\begin{figure}
    \centering
    \includegraphics[width=1.0\linewidth]{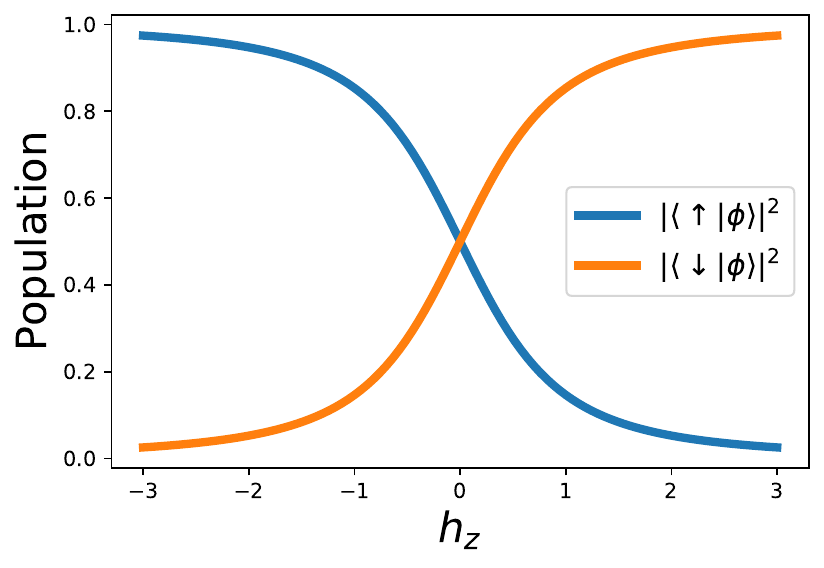}
    \caption{Population of the spin-up state ($\ket{\uparrow}$) and the spin-down state ($|\downarrow\rangle $) after the forward part of the X-rotation gate against $h_{z}$. Here, we adopt the perturbative analysis, and we assume that the initial state is the ground state of the transverse magnetic field$\ket{+}$.
    We can see that almost all angles of the X-rotation gate by controlling the value of $h_{z}$.
    Also, we observe that the behavior of this plot qualitatively coincides with the result of numerical simulation in Fig. \ref{fig:forward_reverse_population_2000} and experimental simulation using a D-Wave device in Fig. \ref{fig:sqr_1qubit_population_hd}.
    }
    \label{fig:perturbation_plot_x_rotation_1qubit}
\end{figure}

We plot the amplitude of the $\ket{\uparrow}$ and $\ket{\downarrow}$ using the equations (\ref{eq:single_x_rotation_analytic_1}) and (\ref{eq:single_x_rotation_analytic_2}) as shown in the Fig \ref{fig:perturbation_plot_x_rotation_1qubit}.

\subsubsection{Perturbative analysis of controlled-not gate operation}

Second, we consider the perturbative analysis of the controlled-not gate.
From the subsection \ref{sec:propose_c-not} in the main text, the degenerated ground states are given by

\begin{align}
    \ket{1}&=\ket{\downarrow\downarrow}\\
    \ket{2}&=\ket{\downarrow\uparrow}.
\end{align}


The drive Hamiltonian is given by the equation (\ref{eq:c-not_drive_ham}).

Also, to discuss general circumstances that include the cases in section \ref{sec:experimental_demonstration} and section \ref{sec:realization_of_gate_op} as special cases, we consider the inhomogeneous longitudinal magnetic field as follows
\begin{align}
    H_{C}=\hat{\sigma}_{1}^{(z)}+\eta\hat{\sigma}_{2}^{(z)}.
\end{align}

We remark that in the case where $\eta=1.0$,  the scenario aligns with the formulation presented in section \ref{sec:realization_of_gate_op} and the numerical simulation detailed in section \ref{sec:numerical_simulation}.
On the other hand, when $\eta=0.3$, the situation corresponds to the experimental results using the D-Wave device discussed in section \ref{sec:realization_of_gate_op}.
In this case, each transition matrix of the perturbation operator $V_{nm}, n,m=1,2$ is given by
\begin{align}
    V_{11}&=-h_{z}(1+\eta)\label{eq:cnot_2qubit_v11}\\
    V_{22}&=-h_{z}(1-\eta)\\
    V_{12}&=-\frac{1}{2}\label{eq:cnot_2qubit_v12}\\
    V_{21}&=-\frac{1}{2}.
\end{align}
We obtain the first-order perturbation energy $E_{0,\pm}^{(1)}$ as follows.


\begin{align}
    E_{0,\pm}^{(1)}=\frac{1}{2}\biggl\{-2h_{z}\pm\sqrt{\bigl(2\eta h_{z}\bigr)^{2}+1}\biggr\}\label{eq:cnot_2qubit_E1}.
\end{align}

Using the (\ref{eq:cnot_2qubit_v11}), (\ref{eq:cnot_2qubit_v12}), (\ref{eq:cnot_2qubit_E1}), and, (\ref{eq:ratio_c1c2}), the ratio $c_{\pm,1}/c_{\pm,2}$ is given by


\begin{align}
    \frac{c_{\pm,1}}{c_{\pm,2}}=2h_{z}\eta\mp\sqrt{(2h_{z}\eta)^{2}+1}.
\end{align}

Finally, from the normalization of the state $c_{\pm,1}^{2}+c_{\pm,2}^{2}=1$, the population of the $c_{\pm,2}^{2}$ is derived as follows.

\begin{align}
    c_{\pm,1}^{2}&=\frac{K_{(\mp)}^{2}}{K_{(\mp)}^{2}+1}\\
    c_{\pm,2}^{2}&=\frac{1}{K_{(\mp)}^{2}+1}
\end{align}

where
\begin{align}
    K_{(\mp)}\equiv 2h_{z}\eta\mp\sqrt{(2h_{z}\eta)^{2}+1}.
\end{align}

\begin{figure}
    \centering
    \includegraphics[width=1\linewidth]{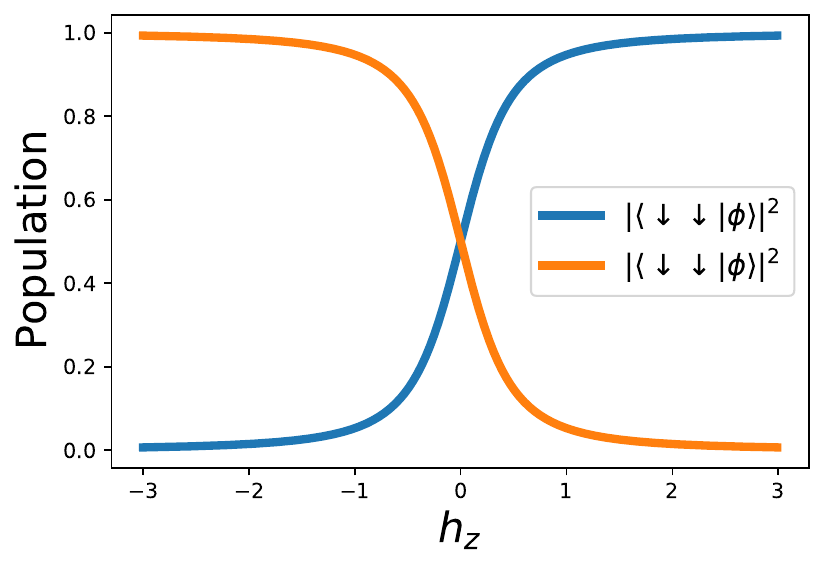}
    \caption{Population of the $\ket{\downarrow\downarrow}$ and $\ket{\downarrow\uparrow}$ after the forward part of the controlled-not gate operation.
    Here, we adopt the perturbative analysis when parameter $\eta=1.0$ in the equation (\ref{eq:c-not_prob_ham}) and we assume that
    the initial state is the ground state of the transverse magnetic field$\ket{++}$.
    It is worth mentioning that this case corresponds to the standard formulation in section.\ref{sec:propose_c-not}.
    We observe almost all angles of the controlled-not gate by controlling the value of $h_z$.
    }
    \label{fig:population_hz_a1}
\end{figure}

\begin{figure}
    \centering
    \includegraphics[width=1\linewidth]{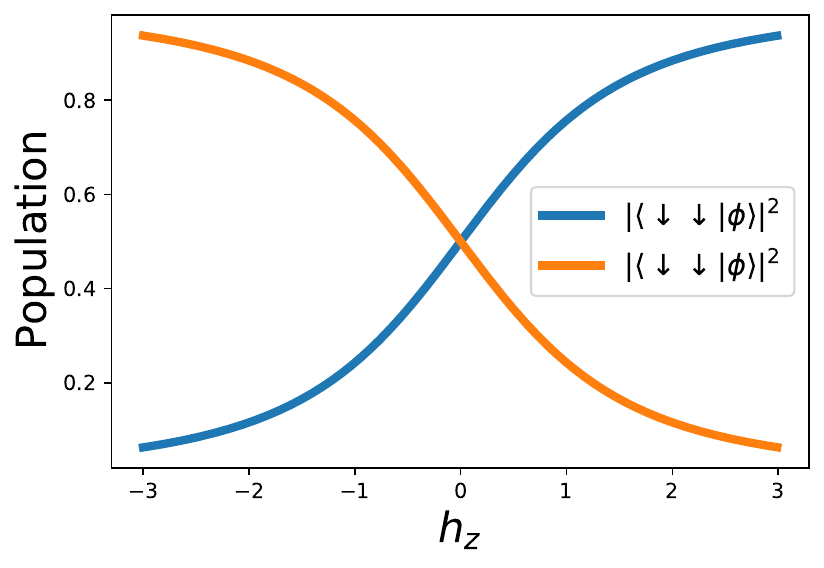}
    \caption{Population of the $\ket{\downarrow\downarrow}$ and $\ket{\downarrow\uparrow}$ after the forward part of the controlled-not gate operation. 
    Here, we use the perturbative analysis when the parameter is $\eta=0.3$ in the equation (\ref{eq:c-not_prob_ham}), and we assume that the initial state is the ground state of the transverse magnetic field$\ket{++}$.
    It is worth mentioning that this case corresponds to the formulation with the parameter used in the D-wave demonstration in section.\ref{sec:c_not_d-wave}.
    We observe almost all angles of the controlled-not gate by controlling the $h_z$.}
    \label{fig:population_hz_a03}
\end{figure}

In the inhomogeneous parameter $\eta=1.0$ we plot the population of the $c_{\pm,1}^{2}$ and $c_{\pm,2}^{2}$ against $h_{z}$ as shown in Fig.\ref{fig:population_hz_a1}.
Also, in the case where the inhomogeneous parameter $\eta=0.3$ the population of the $c_{\pm,1}^{2}$ and $c_{\pm,2}^{2}$ against $h_{z}$ is ploted as shown in Fig.\ref{fig:population_hz_a03}.

\section{Detail of Result for Numerical simulation}\label{sec:detail_numerical_simulation}

This appendix delves into the comprehensive analysis of the numerical
results, elucidating the behaviors observed for both the X-rotation gate and the controlled-not gate operations under varying conditions.

\subsection{Rotation about the X-axis}

\begin{figure*}
    \centering
    \includegraphics[width=180mm]{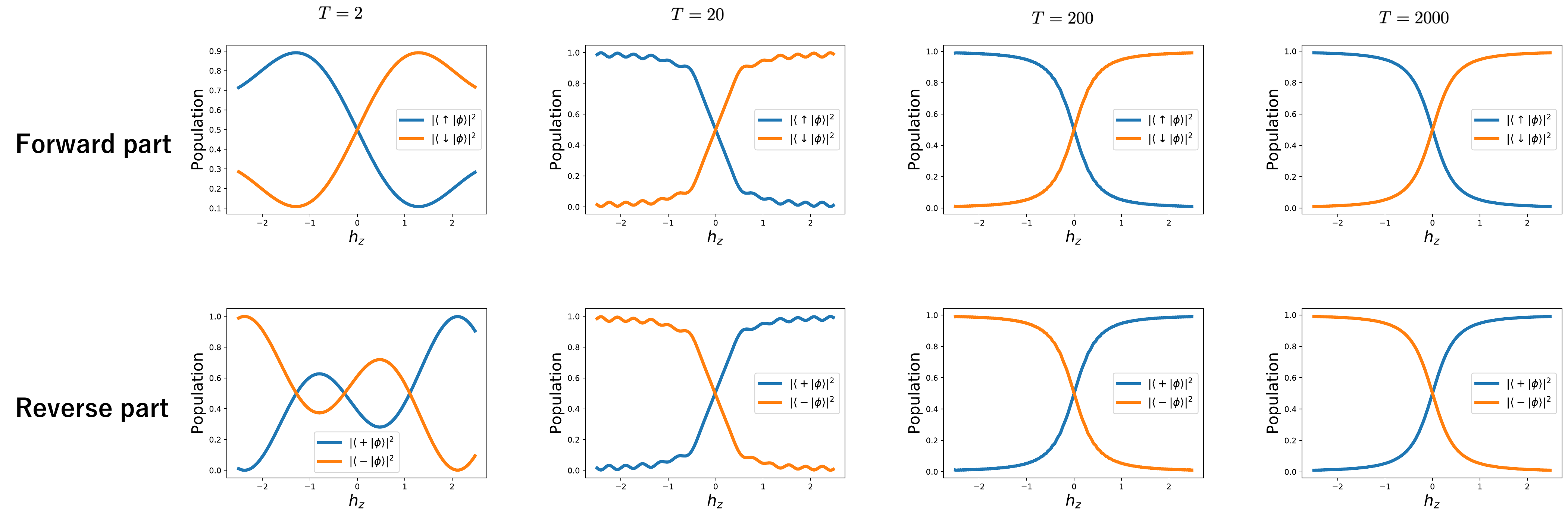}
    \caption{Numerical result of the X-rotation gate after the forward part and the reverse part for several annealing times $T$.
    These plots show the population after the forward part and the reverse part against the amplitude of the longitudinal magnetic field via forward part $h_{z}$.
    We choose the upstate $\ket{+}$ as the initial state.}\label{fig:1logical_1phys_x_rotation_forward_reverse_pop}
\end{figure*}

The results for the  X-rotation gate, as performed in Section \ref{sec:X-axis_gate_numerical_sinulation}, are represented in Fig.\ref{fig:1logical_1phys_x_rotation_forward_reverse_pop}.
Here, we numerically solve the Schr\"{o}dinger equation, and
we plot the population of the spin-up state and spin-down state against $h_z$ after the forward part and reverse part. We confirm that, as we increase $T$, these numerical results converge to the analytical results in Fig.\ref{fig:perturbation_plot_x_rotation_1qubit}.
Therefore, our approach realizes rotation about the X-axis at any desired angle.




\subsection{Controlled-Not gate}

\begin{figure*}
    \centering
    \includegraphics[width=150mm]{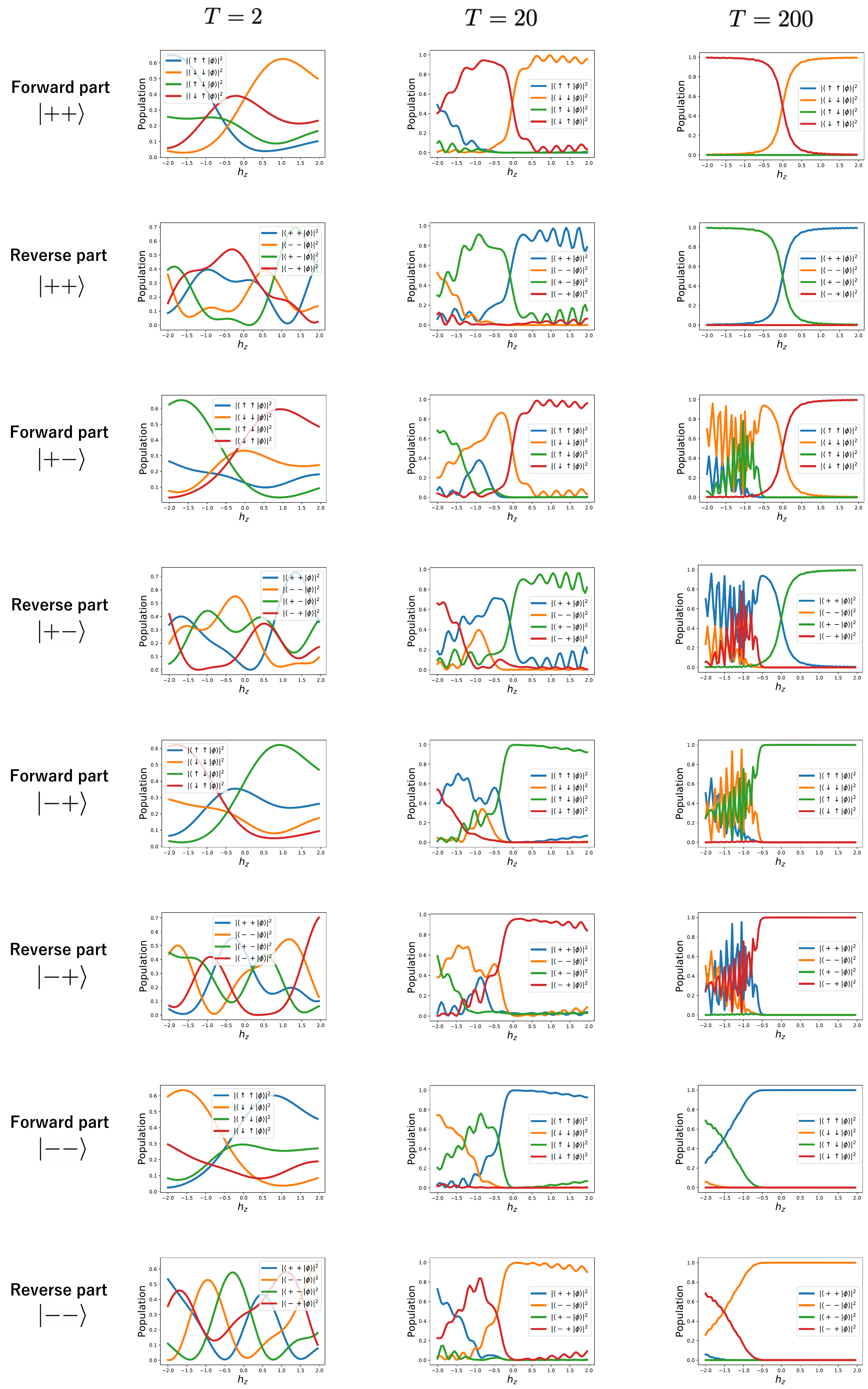}
    \caption{Numerical result of the controlled-not gate after the forward part and the reverse part for annealing time $T=2,20,200$.
    These plots show the population after the forward part and the reverse part against $h_{z}$ for the several initial states.
    }
    \label{fig:cnot-1physical_qubit_detail_numerical1}
\end{figure*}

\begin{figure*}
    \centering
    \includegraphics[width=100mm]{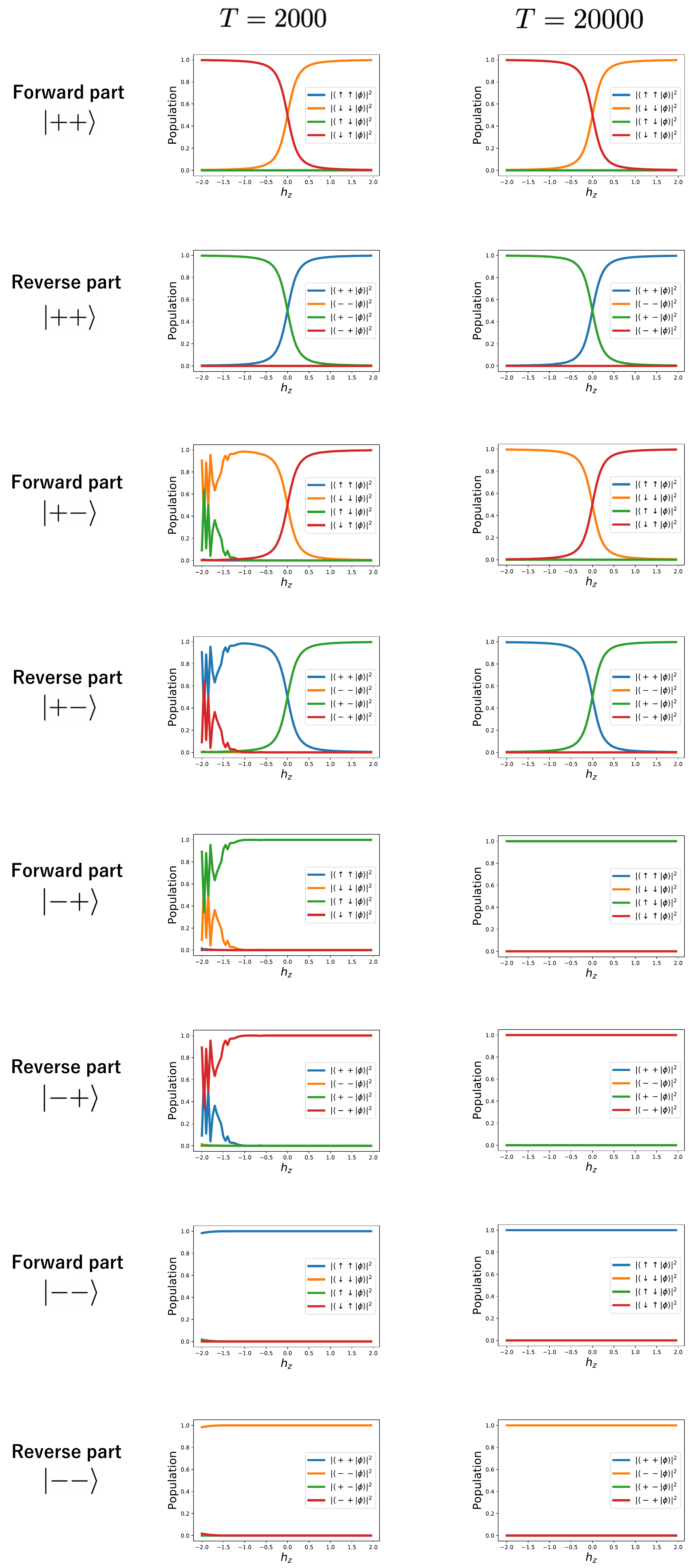}
    \caption{Numerical result of the controlled-not gate after the forward part and the reverse part for annealing time $T=2000,20000$.
    These plots show the population after the forward part and the reverse part against $h_{z}$ for the several initial states.}
    \label{fig:cnot-1physical_qubit_detail_numerical2}
\end{figure*}

We display the results of the numerical simulation for the controlled-not gate in Fig.\ref{fig:cnot-1physical_qubit_detail_numerical1} and Fig.\ref{fig:cnot-1physical_qubit_detail_numerical2} in detail. 
The setup is the same as that in the section.\ref{sec:c_not_d-wave}.
From Fig.\ref{fig:cnot-1physical_qubit_detail_numerical1} and Fig.\ref{fig:cnot-1physical_qubit_detail_numerical2}, we observe that, as we increase $T$, the numerical results approach to the analytical results shown in Fig.\ref{fig:population_hz_a1} for the initial state of $\ket{++}$.
Moreover, by controlling the value of $h_{z}$, we can control the population of the energy eigenstates after the reverse part when the initial state of the control qubit is $|+\rangle $. Therefore, we can implement the control-not gate with our method.

\end{document}